\documentclass[twocolumn,apj,numberedappendix,appendixfloats]{emulateapj}
\usepackage{apjfonts}
\epsscale{1.15}
\newcommand{\rotate}{}

\shorttitle{The Nature of Double-Peaked [O\,{\sc iii}] AGNs} 
\shortauthors{Fu et al.}
\journalinfo{ApJ accepted 2011 October 24}
\submitted{ApJ accepted 2011 October 24}

\newcommand{\chandra}{{\it Chandra}}
\newcommand{\hst}{{\it HST}}
\newcommand{\kms}{{km s$^{-1}$}}
\newcommand{\msun}{$M_{\odot}$}
\newcommand{\lsun}{$L_{\odot}$}
\newcommand{\nd}{\nodata}
\newcommand{\Ha}{H$\alpha$}
\newcommand{\Hb}{H$\beta$}
\newcommand{\OII}{[O\,{\sc ii}]}
\newcommand{\SII}{[S\,{\sc ii}]}
\newcommand{\OIII}{[O\,{\sc iii}]}
\newcommand{\OIIIexpanded}{[O\,{\sc iii}]\,$\lambda$5007}
\newcommand{\NII}{[N\,{\sc ii}]}

\begin{document}

\title{The Nature of Double-Peaked [O III] Active Galactic Nuclei\altaffilmark{*}}

\altaffiltext{*}{Some of the data presented herein were obtained at the W.M. Keck Observatory, which is operated as a scientific partnership among the California Institute of Technology, the University of California and the National Aeronautics and Space Administration. The Observatory was made possible by the generous financial support of the W.M. Keck Foundation.}

\author{Hai Fu\altaffilmark{1}, Lin Yan\altaffilmark{2}, Adam D. Myers\altaffilmark{3,6}, Alan Stockton\altaffilmark{4}, S. G. Djorgovski\altaffilmark{1,7}, G. Aldering\altaffilmark{5}, and Jeffrey A. Rich\altaffilmark{4}
}
\altaffiltext{1}{Astronomy Department, California Institute of Technology, MS 249$-$17, Pasadena, CA 91125, USA; fu@astro.caltech.edu} 
\altaffiltext{2}{Spitzer Science Center, California Institute of Technology, MS 220$-$06, Pasadena, CA 91125, USA}  
\altaffiltext{3}{Department of Physics and Astronomy, University of Wyoming, Laramie, WY 82071, USA}
\altaffiltext{4}{Institute for Astronomy, University of Hawaii, 2680 Woodlawn Drive,
Honolulu, HI 96822, USA}
\altaffiltext{5}{Physics Division, Lawrence Berkeley National Laboratory, 1 Cyclotron Road, Berkeley, CA 94720, USA}
\altaffiltext{6}{Max-Planck-Institut f\"ur Astronomie, K\"onigstuhl 17, D-69117 Heidelberg, Germany}
\altaffiltext{7}{Distinguished Visiting Professor, King Abdulaziz University, Jeddah, Saudi Arabia}

\begin{abstract} 
Active galactic nuclei (AGNs) with double-peaked \OIII\ lines are suspected to be sub-kpc or kpc-scale binary AGNs. However, pure gas kinematics can produce the same double-peaked line profile in spatially integrated spectra. Here we combine integral-field spectroscopy and high-resolution imaging of 42 double-peaked \OIII\ AGNs from the Sloan Digital Sky Survey to investigate the constituents of the population. We find two binary AGNs where the line-splitting is driven by the orbital motion of the merging nuclei. Such objects account for only $\sim$2\% of the double-peaked AGNs. Almost all ($\sim$98\%) of the double-peaked AGNs were selected because of gas kinematics; and half of those show spatially resolved narrow-line regions that extend 4$-$20 kpc from the nuclei. Serendipitously, we find two spectrally {\em unresolved} binary AGNs where gas kinematics produced the double-peaked \OIII\ lines. The relatively frequent serendipitous discoveries indicate that only $\sim$1\% of binary AGNs would appear double-peaked in Sloan spectra and $2.2_{-0.8}^{+2.5}$\% of all Sloan AGNs are binary AGNs. Therefore, the double-peaked sample does not offer much advantage over any other AGN samples in finding binary AGNs. The binary AGN fraction implies an elevated AGN duty cycle ($8_{-3}^{+8}$\%), suggesting galaxy interactions enhance nuclear accretion. We illustrate that integral-field spectroscopy is crucial for identifying binary AGNs: several objects previously classified as ``binary AGNs" with long-slit spectra are most likely single AGNs with extended narrow-line regions. The formation of extended narrow-line regions driven by radiation pressure is also discussed.
\end{abstract}


\section{Introduction} \label{sec:introduction}

The ubiquity of binary supermassive black holes (SMBHs) is expected because of two commonly accepted facts \citep{Begelman80}: that galaxies frequently merge \citep{Toomre72} and that SMBHs occupy the centers of the majority of massive galaxies \citep{Kormendy95,Richstone98}. If mergers trigger or enhance nuclear accretion as predicted by simulations \citep{Barnes96}, a significant fraction of mergers should appear as binary active galactic nuclei (AGNs). 

Over the last three decades, binary AGNs have been found with separations ranging from parsec to tens of kiloparsec (kpc) scales. Thanks to the Sloan Digital Sky Survey (SDSS), we now have a large sample of binary quasars with separations of tens of kpcs; and they account for $\sim$0.1\% of the QSO population at $z > 1$ \citep{Djorgovski87, Djorgovski91, Kochanek99, Hennawi06, Myers07, Myers08, Green10} and $\sim$3.6\% of SDSS spectroscopically selected AGNs between $0.02 < z < 0.16$ \citep{Liu11a}. On kpc-scales, we have six convincing examples, LBQS 0103-2753 \citep{Junkkarinen01,Shields11}, NGC~6240 \citep{Komossa03}, 3C~294 \citep{Stockton04}, Mrk~463 \citep{Bianchi08}, Mrk~739 \citep{Koss11}, and SDSS\,J150243.1$+$111557 \citep{Fu11c}. On parsec-scales, there is only one convincing example, the radio galaxy 0402+379, for which only the Very Long Baseline Array (VLBA) can resolve the 7-parsec-separation binary \citep{Rodriguez06}. 

Serendipitously discovered binary AGNs are of course interesting objects to study, but in order to understand the effect of galaxy interactions on nuclear activities, it is crucial to estimate the frequency of binary AGNs as a function of physical separation. Systematic searches for parsec-scale binary AGNs have focused on objects with two distinct sets of AGN emission lines, either as double-peaked broad emission lines \citep{Gaskell83,Gaskell96,Boroson09} or as narrow lines offset from the broad lines \citep{Komossa08,Eracleous11,Tsalmantza11}. However, double-peaked broad lines are not necessarily binary AGNs---they could be equally well explained as a Keplerian accretion disk \citep{Chen89,Eracleous97,Eracleous03}. And the narrow-line offset systems could be gravitational wave recoiled black holes \citep{Komossa08}. Unfortunately, it is difficult to confirm or refute these binary AGN candidates, because of the extremely high spatial resolution needed to resolve the components and/or the long time span needed to monitor the systematic velocity changes due to orbital motion.

The same technique has been used to find binary AGNs with separations larger than the scale of the broad-line regions but within $\sim$10 kpcs. Several systematic searches have compiled AGNs with double-peaked \OIIIexpanded\ emission lines (dpAGNs hereafter) from the DEEP2 survey \citep{Gerke07,Comerford09a} and the SDSS \citep{Xu09,Wang09,Liu10a,Smith10,Liu11a}. If double-peaked \OIII\ emission is indicative of a binary AGN, such emission most likely represents objects with transverse separations of $\sim$100 parsecs to $\sim$10 kpcs \citep[e.g.,][]{Wang09}. However, double-peaked emission lines may also arise from mechanisms other than orbital motion, such as gas kinematics in the narrow-line regions \citep[e.g.,][]{Gelderman94,Fu09,Fischer11,Shen11}, or jet-cloud interactions \citep[e.g.,][]{Stockton07, Rosario10}.

Unlike parsec-scale binaries, it is relatively easy to confirm or refute the kpc-scale binary AGN candidates with existing facilities. \citet{Shen11} have recently carried out an imaging and long-slit spectroscopic study of 31 SDSS dpAGNs (mostly at declinations of $< 22^\circ$). They identified five binary AGNs, with each resolved galaxy component associated with AGN-photoionized narrow emission lines (i.e., type-2$-$type-2 pairs). However, their study was limited by the natural seeing and the spatial coverage of the long-slit. To move forward, we decided to carry out a high-resolution imaging and integral-field spectroscopic survey of dpAGNs.

Ground-based imaging aided by adaptive optics enables efficient high-spatial-resolution imaging surveys. In \citet{Fu11a}, we published images of 50 SDSS dpAGNs from the Keck laser guide star adaptive optics system \citep[LGSAO;][]{Wizinowich06}. With a spatial resolution of $\sim$0.1\arcsec, we found $\sim$30\% of dpAGNs at $z < 0.6$ show discernible companions. Using the same instrument, \citet{Rosario11} found a similar result with a smaller sample. These results showed that at least 70\% of the dpAGNs are either single AGNs or binary type-2 AGNs with coalesced host galaxies. It is worth noting that even a merging galaxy associated with a dpAGN could be ascribed to single AGNs---the double-peaked AGN emission lines could arise from a single active component which is merging with a non-active galaxy.

We have extended our LGSAO imaging to a sample doubling the size of that surveyed by \citet{Fu11a}. In addition, we have obtained integral-field spectroscopy of 42 dpAGNs, including most of the merging systems identified with LGSAO imaging. In this paper, we analyze the combined data set to identify kpc-scale binary AGNs and to study the statistics of dpAGN types. We describe our observations in Section~\ref{sec:obs}. In Section~\ref{sec:class}, we classify the AGNs based on both stellar morphology from high-resolution imaging and ionized gas morphology and kinematics from integral-field spectroscopy. In Section~\ref{sec:comparison} we compare the classifications based on previous long-slit spectroscopy and our integral-field spectroscopy. We show that several objects classified in the literature as ``binary" are, most likely, single AGNs with extended narrow-line regions. We conclude by discussing the binary AGN fraction, the efficiency of the double-peaked selection technique, and the formation of extended narrow-line regions (Section~\ref{sec:discuss}).

Throughout we adopt a $\Lambda$CDM cosmology with $\Omega_{\rm m}=0.3$, $\Omega_\Lambda=0.7$ and $h \equiv$ H$_0$/100 km~s$^{-1}$~Mpc$^{-1}$ = 0.7, consistent with the maximum likelihood estimates from WMAP \citep{Dunkley09}. 

\section{Observations} \label{sec:obs}

The dpAGNs presented here were selected from the SDSS DR7 spectroscopic data set by several groups \citep[][]{Wang09,Liu10a,Smith10}. There are a total of 340 dpAGNs between $0.008 < z < 0.686$, or $\sim$1\% of the entire SDSS AGN sample. Their \OIIIexpanded\ lines show double-peaked profiles with velocity splitting ranging from 151 to 1314 \kms. Initially, it was suspected that most of these double-peaked line profiles could originate from the orbital motions of binary AGNs blended within the 3\arcsec-diameter apertures of SDSS fibers \citep[e.g.,][]{Wang09}. Nonetheless, to test whether these are binary AGNs, the SDSS data are insufficient because of their limited spatial resolution. To better understand this population of AGNs, we obtained high-resolution ($\sim$0.1\arcsec) adaptive-optics imaging and integral-field spectroscopy for a representative sample of dpAGNs. 

\subsection{High Resolution Imaging} \label{sec:nirc2}

Six of the dpAGNs have ACS/WFPC2 images in the \hst\ archive (SDSS\,J0807+3900, SDSS\,J0941+3944, SDSS\,J1129+5756, SDSS\,J1301-0058, SDSS\,J1526+4140\footnote{SDSS\,J1526+4140 is also known as NGC\,5929} and SDSS\,J2310-0900). For the rest, we attempted high resolution near-infrared (NIR) imaging with the Keck\,II laser guide-star adaptive-optics system \citep[LGSAO;][]{Wizinowich06}. LGSAO observations require a bright ($R \lesssim 18$) star within $\sim$60\arcsec\ of the target for tip-tilt corrections. Therefore, we selected 45 type-1 (broad-line) AGNs and 107 type-2 (narrow-line) AGNs that have nearby tip-tilt stars. We further included 23 type-1 AGNs without nearby tip-tilt stars but brighter than $R = 18$ so that the AGN itself could serve as the beacon for tip-tilt corrections. Our final LGSAO target sample contained 173 AGNs over the redshift range $0.023 < z < 0.686$.

We obtained $H$-band (1.473--1.803 $\mu$m; OSIRIS/Hbb filter) images for three sources (SDSS\,J0400-0652, SDSS\,J0952+2552, and SDSS\,J1240+3534) with the OSIRIS imager \citep{Larkin06} at 0.02\arcsec\ pixel$^{-1}$ on 2010 March 6 and 7 (UT), and $K^{\prime}$-band (or $K_{\rm p}$, 1.948--2.299 $\mu$m) images for 97 sources on 2010 June 3 and 4 (UT) and 2011 January 5 and 6 (UT) with NIRC2 at 0.04\arcsec\ pixel$^{-1}$. For each target, we took at least three 1-minute exposures, dithered within boxes of 4\arcsec\ to 7\arcsec\ width. For the brightest targets, we used shorter exposure times and a number of co-adds to avoid saturation. The final exposure times and the CFHT/DIMM seeing data are listed in Tables~\ref{tab:imglog_merger} \& \ref{tab:imglog_single} in the Appendix. Half of our imaging data were presented in \citet{Fu11a} and we followed the same data reduction procedure here. The astrometry of these images are tied to that of the SDSS using the centroid position of the targets and assuming the pre-determined instrument plate scale.

In summary, we have high-resolution images (FWHM [Full Width Half Maximum] $\sim$ 0.1\arcsec) for a total of 106 dpAGNs --- 37 type-1 and 63 type-2 AGNs were imaged by the Keck\,II LGSAO in the NIR, and 6 type-2 AGNs were imaged by the \hst\ in the optical. Inspecting the images, we found that 31 ($29^{+5}_{-4}$\%)\footnote{We estimate the 1$\sigma$ binomial confidence intervals of any fractions from the quantiles of the beta distribution \citep{Cameron11}.} sources have companions within 3\arcsec\ (Fig.~\ref{fig:imglog_merger}), consistent with the merger fractions found by \citet{Fu11a} and \citet{Rosario11}. 

\subsection{Integral-Field Spectroscopy}

Integral-field spectroscopy allows us to spatially resolve emission lines---thus comparing emission-line morphologies with continuum morphologies---and to derive the kinematics of the emission lines. Based on this information, we can study the constituents of dpAGNs, as it is known that several mechanisms can produce the same double-peaked line profiles \citep[e.g.,][]{Rosario10,Fu11a,Shen11}. The sample that we compiled for integral-field spectroscopy is biased in favor of sources with existing high-resolution imaging data and sources that displayed close companions within 3\arcsec. We obtained seeing-limited optical integral-field spectroscopy for 39 sources and LGSAO-aided integral-field spectroscopy for 3 additional sources. The combined spectroscopic sample includes 26 apparent mergers that show close companions within 3\arcsec, 11 isolated sources, and 5 sources with only seeing-limited images (Table~\ref{tab:ifulog}). We assume that the 5 sources without high-resolution images are single galaxies.

\subsubsection{Adaptive Optics-Aided Integral-Field Spectroscopy} \label{sec:osiris}

We obtained NIR integral-field spectra with the OH-Suppressing Infrared Imaging Spectrograph \citep[OSIRIS][]{Larkin06} on the Keck\,II telescope for 3 double-peaked AGNs (SDSS\,J0808+4813, SDSS\,J1050+0839, and SDSS\,J1240+3534). OSIRIS was operated with the LGSAO system to deliver high spatial-resolution datacubes.

The OSIRIS observations took place on 2010 March 7 (UT) under excellent conditions. We used the 0.05\arcsec\ lenslet scale and either the Kn2 (2.036$-$2.141\,$\mu$m) or the Kn3 (2.121$-$2.229\,$\mu$m) filter to cover the redshifted Pa$\alpha$\,$\lambda1.87\,\mu$m line with a 2.25\arcsec$\times$3.2\arcsec\ (Kn2) or 2.4\arcsec$\times$3.2\arcsec\ (Kn3) field of view and a spectral resolution $R \simeq 3800$. Three or four on-source exposures of 600 seconds were taken with a 600-second sky frame in between. An A1V star, HIP\,49198, was observed for telluric absorption correction and flux calibration.

We reduced the data using the OSIRIS data reduction pipeline \citep[version 2.3;][]{Krabbe04}. In summary, the pipeline subtracts darks and biases from the raw images, removes crosstalk signals, identifies glitches, cleans cosmic-rays, extracts and wavelength-calibrates spectra from individual lenslets, assembles the datacubes, corrects for differential atmospheric refraction, subtracts sky using an implementation of the scaled sky subtraction algorithm \citep{Davies07b}, removes telluric absorption, and co-adds the individual exposures to form the final datacube. The spatial resolution of the final datacubes is $\sim$0.12\arcsec.

\subsubsection{Seeing-Limited Integral-Field Spectroscopy} \label{sec:snifs}

We obtained seeing-limited integral-field spectra using the Supernova Integral-Field Spectrograph \citep[SNIFS;][]{Aldering02,Lantz04} on the University of Hawaii 2.2-meter telescope (Mauna Kea) in a fully remote observing mode. SNIFS is a pure-lenslet integral-field spectrograph similar to TIGER \citep{Bacon95}. It has a fully filled 6.45\arcsec$\times$6.45\arcsec\ field of view sampled by a 15$\times$15 grid of 0.43\arcsec$\times$0.43\arcsec\ spatial elements. The dual-channel spectrograph covers wavelengths from 3200 to 10000\,\AA\ simultaneously with a spectral resolution of FWHM = 6.1 and 8.2\,\AA\ for the blue (3200$-$6000\,\AA; $R \sim 700$) and red (5000$-$10000\,\AA; $R \sim 1000$) spectrographs, respectively. The instrument position angles (PAs) are fixed on the sky --- $2.6\pm0.8^{\circ}$ for the blue and $5.0\pm1.2^{\circ}$ for the red. We measured the spectral resolution from arc lamp exposures taken with each science exposure and they agree well with the measurements from the night sky lines in the final reduced datacubes. 

Our SNIFS observations took place on the first half-nights of 2010 August 4, 6, 8, 9 and on the two full nights of 2011 March 26, 29 (UT). In addition, a 15-minute exposure of SDSS\,J1155+1507 was taken on 2010 November 21 (UT). The 2010 run was photometric with seeing between 0.6\arcsec\ and 1.2\arcsec, and the 2011 run was cloudy with seeing between 0.8\arcsec\ and 2\arcsec. Depending on the [O\,{\sc iii}]\,$\lambda$5007 line fluxes and the observing conditions, the total integration time ranged from 10 to 60 minutes for each object. Table~\ref{tab:ifulog} presents the observing log.

The raw data were first passed through the SNIFS data reduction pipeline \citep{Aldering06}. After subtraction of biases and a diffuse-light component, the pipeline extracted the spectra from the CCD and reassembled them into ($x, y, \lambda$)-datacubes, which were subsequently wavelength calibrated, flat-fielded, and cosmic-ray cleaned. Then, we used our custom IDL scripts to subtract the sky using regions free of object light, to rectify the field to PA = 0, to correct for differential atmospheric refraction as computed by the SLA\_REFRO routine in the SLALIB library\footnote{http://star-www.rl.ac.uk/star/docs/sun67.htx/sun67.html} (version 2.5-4), and to join the blue and red channel datacubes. For flux calibration, we used the SDSS spectra. We used standard star observations exposed on the same nights as the data frames to remove telluric absorption features. Finally, Galactic extinction was removed using the dust map of \citet{Schlegel98} and the standard reddening curve of \citet{Cardelli89} with $R_V$ = 3.1. Considering only spectra with \OIIIexpanded\ peak-to-noise ratio greater than 2.0 as valid, we reached a surface brightness depth of $1.4\pm1.0\times10^{-16}$\,erg\,s$^{-1}$\,cm$^{-2}$\,arcsec$^{-2}$ (Table~\ref{tab:ifulog} Column 6). The depth varies with exposure time, observed wavelength, and atmosphere conditions. But for all of the sources, our datacubes are sensitive enough to detect both \OIII\ components seen in the SDSS spectra with at least five spatial elements, assuming the surface brightness profile of the components follows a Gaussian distribution with FWHM less than 2\arcsec. We also confirm that SNIFS detects both \OIII\ components by comparing the \OIII\ line profiles extracted from the datacubes with those from SDSS smoothed to SNIFS resolution. 

We spatially register the datacubes with the high-resolution images by measuring the centroid position of the target from emission-line-free continuum images stacked from the datacubes. When there are multiple components, we use the centroid of the brightest component. For unresolved sources, the centroid accuracy is approximately equal to FWHM/SNR (signal-noise-ratio). The stacked continuum images from the SNIFS datacube have rather high SNR, so the centroids can be pinned down better than 0.05" given the typical resolution of FWHM = 1.0", despite that the pixel size is 0.4". The centroids from the Keck and \hst\ images can be measured more accurately than the SNIFS images.

The instrument PA and plate scale previously determined by the SNIFS instrument team are verified by comparing the observed differential atmospheric refraction from standard stars with that predicted by SLA\_REFRO using the observed atmosphere conditions. 

\section{Classification} \label{sec:class}

\begin{figure*}[!t]
\epsscale{0.87}
\plotone{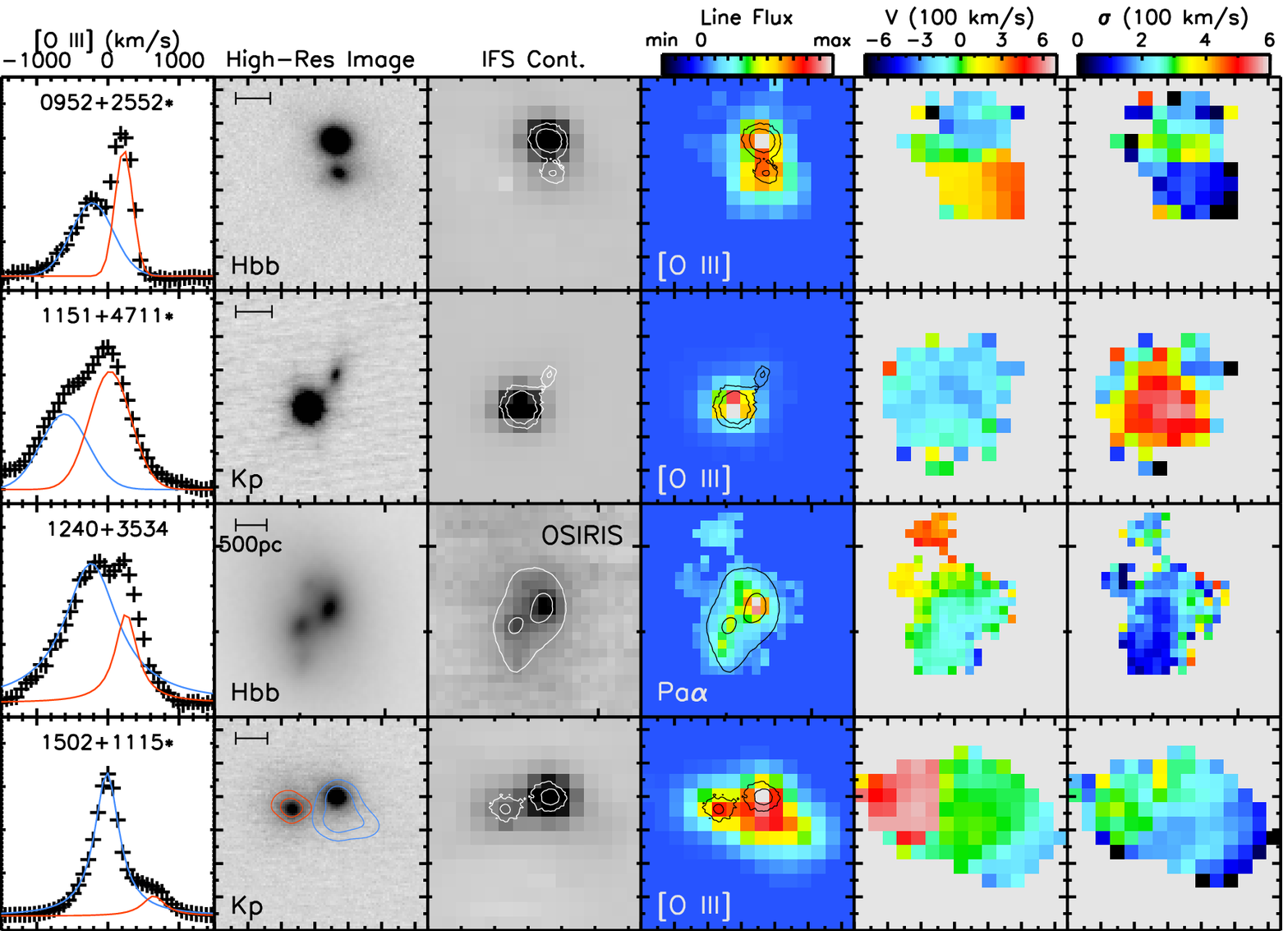}
\epsscale{1.0}
\caption{Binary AGNs. For each object, from left to right are: the SDSS \OIII\,$\lambda$5007 line profile, the highest-resolution broad-band image, the emission-line--free continuum from the datacube, the emission-line intensity map, the velocity map, and the velocity dispersion map. Object designations are labeled in the first column and broad-line AGNs are indicated with stars. The two-component fit to the \OIII\ line profile is overplotted. The filter for the broad-band images are labeled, and the scale bar indicates a transverse separation of 5~kpc, unless otherwise indicated. For objects where SNIFS spectrally resolves the kinematic components, we overlay the emission-line fluxes from the blueshifted and redshifted components as blue and red contours on the broad-band image. Contours in third and fourth columns are from the broad-band images in the second column, which have been spatially aligned with the datacubes. The emission line that we measured are labeled in the fourth column. In all of the images, N is up and E is to the left; major tickmarks are spaced in 1\arcsec. Note that the maps of SDSS\,J1240+3534 are from OSIRIS.
\label{fig:binary}} 
\end{figure*}

\begin{figure}[!t]
\plotone{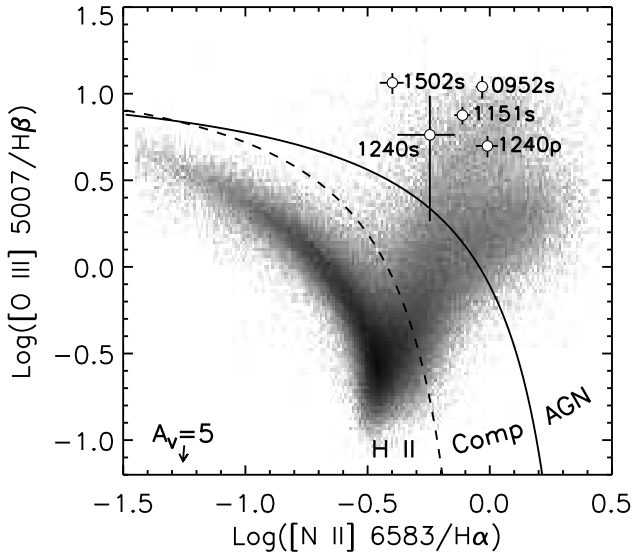}
\caption{BPT diagnostic diagram. The line ratios of the narrow-line components in the binary AGNs indicate AGN photoionization. Our measurements are shown as circles with 1$\sigma$ error bars. For SDSS\,J0952+2552, SDSS\,J1151+4711, and SDSS\,J1502+1115, we only show line ratios of the secondaries (``s") because their primaries (``p") are broad-line AGNs. The background image shows the density distribution of the SDSS emission-line galaxies in log scale. Objects above the solid curve are dominated by AGNs (LINERs are concentrated in the lower denser branch and Seyferts in the upper branch; \citealt{Kewley06}), below the dashed curve are star-forming galaxies \citep{Kauffmann03}, and AGN/star-forming composite galaxies are in between. The arrow shows a reddening vector for a $V$-band extinction of 5 magnitudes.
\label{fig:bpt}}
\end{figure}

\begin{figure}[!t]
\plotone{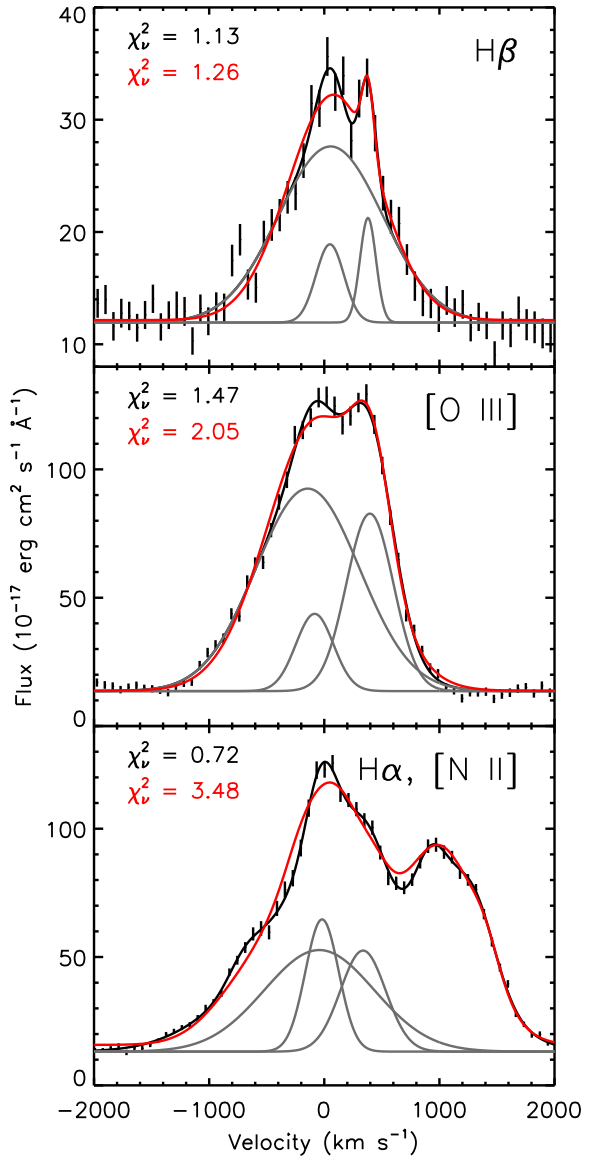}
\caption{Decomposing the emission lines of SDSS\,J1240+3534. The SDSS spectrum are black data points with error bars. The best-fit components from the three-Gaussian models are in grey, and the sum of the components are in black. The broad and narrow blueshifted components are from the primary and secondary nucleus, respectively, and the redshifted component is from the nebula 0\farcs6 NNE of the nuclei (Fig.~\ref{fig:binary}). The best-fit two-Gaussian models are the red curves. The reduced $\chi^2$ values are labeled for the three-Gaussian model and two-Gaussian model in black and red, respectively. 
\label{fig:lineprof}}
\end{figure}

\begin{figure*}[!t]
\epsscale{0.87}
\plotone{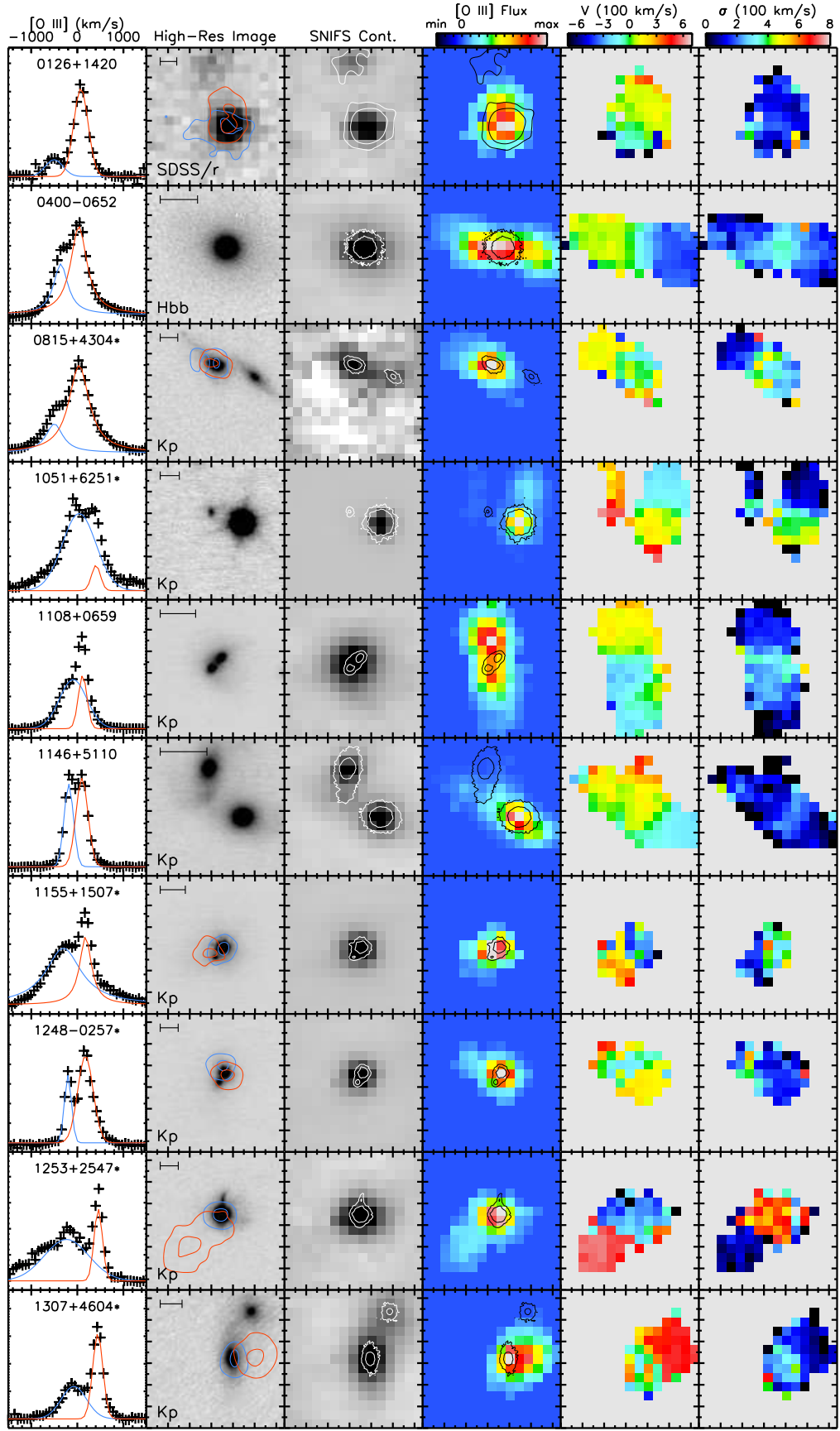}
\caption{Same as Fig.~\ref{fig:binary} but for dpAGNs where extended narrow-line regions produced the double-peaked \OIII\ lines. See Fig.~\ref{fig:binary} for SDSS\,J1240+3534.
\label{fig:enlr}} 
\end{figure*}
\addtocounter{figure}{-1}
\begin{figure*}
\epsscale{0.87}
\plotone{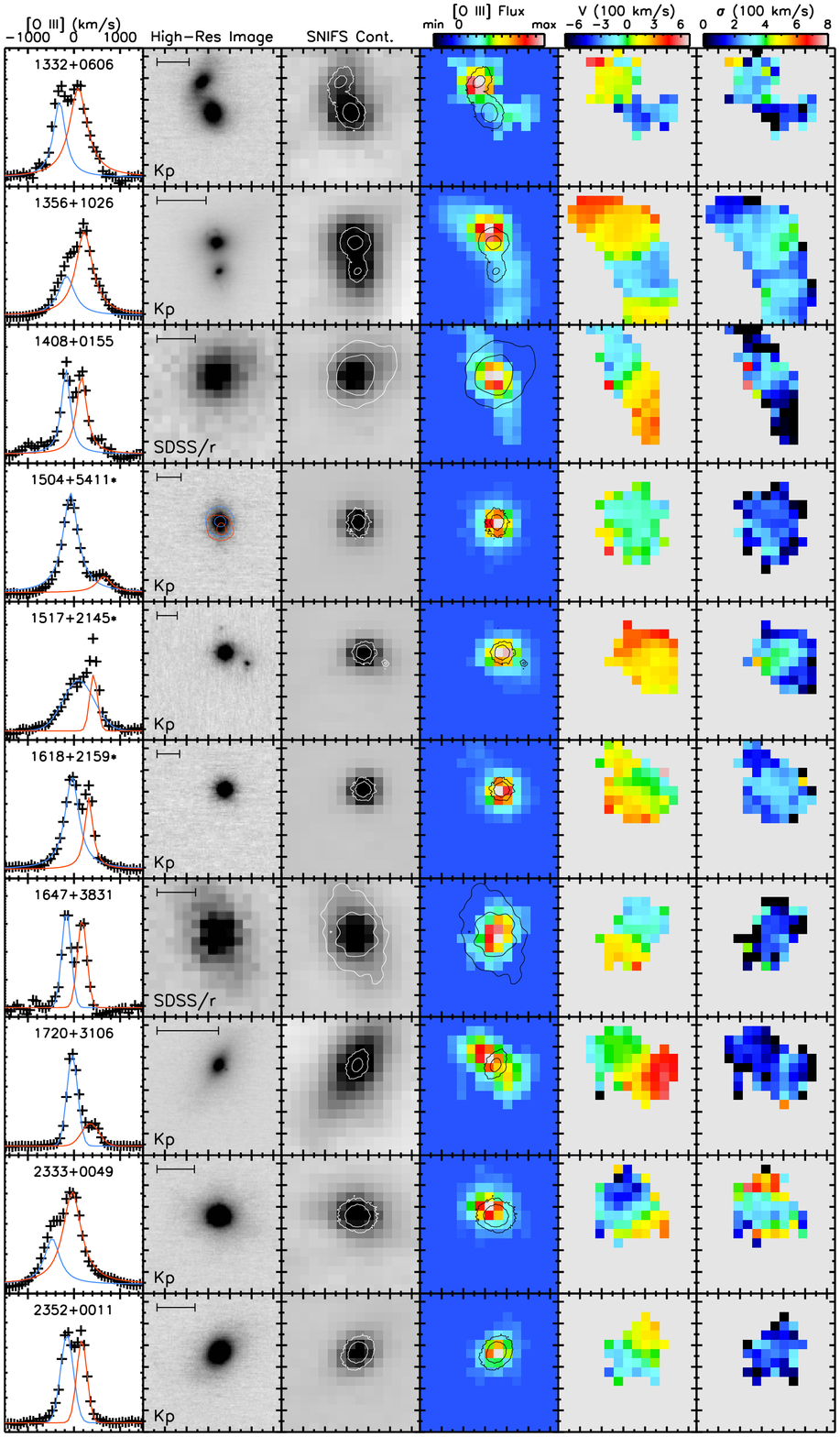}
\epsscale{1.0}
\caption{{\it continued}.}
\end{figure*}

Broadly speaking, the double-peaked \OIII\ line profiles could arise from: (1) the orbital motion of a binary AGN, (2) gas kinematics in extended narrow-line regions (ENLRs), and (3) unresolved nuclear gas kinematics (e.g., aligned outflows or disk rotation on small scales). We classified the 42 IFS-observed dpAGNs into these three broad categories by examining the continuum and emission-line morphologies and the kinematics of the ionized gas. We emphasize that we obtained $\lesssim0.1$\arcsec\ resolution images (\S~\ref{sec:nirc2}) for 37 of the 42 IFS-observed sources.  

We derived maps of emission-line fluxes and kinematics by fitting Gaussians at each spatial element of the datacube. These maps are shown in Figures~\ref{fig:binary}, \ref{fig:enlr}, \& \ref{fig:unresolved} along with the best-available broad-band images. If SNIFS spectrally resolves the double-peaked emission lines, we fit two Gaussians with relative velocities fixed to that determined from the Sloan spectra. In the kinematic maps, we show only the brighter component at each position; but in the line flux maps, we use the total flux. For the OSIRIS data, we fit only single Gaussians because OSIRIS spatially resolved the kinematic components. 

Table~\ref{tab:ifu_sample} summarizes the basic properties of the sample, grouped by our best-effort classification. The reality is, however, much more complicated than this single-parameter classification because of the following facts:

(1) Each object may contain more than two emission-line components. For example, in SDSS\,J1440+6156 the two major kinematic components originate from an unresolved nuclear region, but we clearly detected \OIII\ emission extending $\sim$2\arcsec\ to the north of the nucleus. Our first-order classification is based on the main mechanism that produced the double-peaked \OIII\ line in the Sloan spectra, but we note the second-order mechanisms in Column 11 of Table~\ref{tab:ifu_sample}. 

(2) SNIFS data have lower spatial resolution than that of the AO/\hst\ images. The seeing-limited SNIFS data cannot spatially resolve structures on scales smaller than $\sim$1.2\arcsec. But seven of our SNIFS-observed sources show multiple components within 1.2\arcsec. Except for SDSS\,J0952+2552, which is a clear binary AGN (Fig.~\ref{fig:binary}), they are all difficult to classify. For example, in SDSS\,J1108+0659 the merging galaxies seen in the high-resolution images are completely enshrouded by the ENLR. Although it is clear that the ENLR produces the double-peaked line profile, it remains ambiguous whether both of the merging galaxies host AGNs or a single AGN is powering the ENLR.  Decomposing the velocity components helps the classification, because the flux centroids of each decomposed component can be localized much more accurately than the seeing disk (to $\sim$ FWHM/SNR). But it is impossible to rule out the existence of emission-line regions that spatially coincide with the nuclei. Classifications are thus ambiguous for these six compact mergers. On the other hand, spectral decomposition can only be done for sources where SNIFS resolves the line profiles, which leads us to our third caveat...

(3) ... SNIFS data have lower spectral resolution than that of the SDSS spectra, so it can only spectrally resolve the double-peaked emission-lines when the velocity splitting is sufficiently large and/or the components are sufficiently narrow. 

Bearing these caveats in mind, in the following subsections we describe the three classes of dpAGNs in detail.

\subsection{Binary AGNs} \label{sec:binary}

We found four binary AGNs; these are merging galaxy pairs where the stellar nuclei spatially coincide with distinctive \OIII\ emission-line components (Figure~\ref{fig:binary}). However, there are only two systems, SDSS\,J0952+2552 and SDSS\,1502+1115, whose double-peaked \OIII\ lines are produced by the relative velocities of the merging galaxies. SDSS\,J0952+2552 was also identified as a binary AGN by OSIRIS \citep{McGurk11}. The other two are serendipitous binaries, where ``serendipity" means that the binaries cannot be resolved in velocity space with a spatially integrated spectrum. SDSS\,1240+3534 is double-peaked because of a high-velocity nebula that is spatially offset from the nuclei. The projected separation of the nuclei is only 500~pc. In SDSS\,1151+4711, the spectrally resolved emission-line components are spatially unresolved. Hence, we grouped the last two systems under ``Extended Narrow-Line Region (ENLR)" and ``Unresolved" in Table~\ref{tab:ifu_sample}. We note that the emission lines from the NW companion of SDSS\,1151+4711 are {\em not} scattered light from the primary because they are much narrower.

The primaries\footnote{We define the primaries as the brighter nucleus in the broad-band images.} of three of the four systems are broad-line AGNs (SDSS\,J0952+2552, SDSS\,J1151+4711, and SDSS\,J1502+1115). To test whether their companions are also AGNs, we extracted spectra at the companion locations with 0.85\arcsec\ apertures from the SNIFS datacubes. In all three cases, only narrow lines are seen, and their \OIII/\Hb\ and \NII/\Ha\ ratios place them securely in the Seyfert regime on the classic ``BPT'' \citep{Baldwin81} line-ratio diagnostic diagram (Fig.~\ref{fig:bpt}). Therefore, these are type-1$-$type-2 binaries.

For the narrow-line AGN SDSS\,J1240+3534, the Pa$\alpha$ line is spatially resolved into three major components by OSIRIS (two associated with the nuclei, one offset to the NNE). The velocity map implies that the emission-line components at the two stellar nuclei have very similar line-of-sight velocities. To place them on the BPT diagram, we decompose each of the emission lines by fitting three Gaussians simultaneously with MPFIT \citep{Markwardt09}. We set their initial velocities at those determined from the OSIRIS datacube but allow them to vary during the fit. Figure~\ref{fig:lineprof} shows the result. The three-Gaussian models provide better fits to the SDSS line profiles than two-Gaussian models, with the $\chi^2$ values reduced by 10\%, 30\% and 80\% for \Hb, \OIIIexpanded\ and the \NII$-$\Ha\ complex, respectively. The 1$\sigma$ errors of the parameters were computed from the covariance matrix by MPFIT. The decomposed line ratios of the two nuclear components are consistent with AGN photoionization (Fig.~\ref{fig:bpt}), indicating that SDSS\,J1240+3534 is a type-2$-$type-2 binary.

It should be emphasized that the proof that the gas is photoionized by an AGN does not imply that there is an AGN at the location of the nebula. Because of the proximity between the nuclei in all of these systems, it remains ambiguous whether there is one or two ionizing sources (i.e., AGNs) in each system without high-resolution X-ray or radio images. Recently, the binary nature of SDSS\,J1502+1115 was confirmed by high-resolution radio continuum images from the Expanded Very Large Array (EVLA) \citep{Fu11c}. 

\subsection{Extended Narrow-Line Regions} \label{sec:enlr}

Our IFS data spatially resolved the emission-line nebulae in half (21/42) of the dpAGNs. We refer to such less luminous nebulosity on kpc scales as ENLRs and reserve the term \emph{Extended Emission-Line Region} (EELR) for luminous nebulae on tens of kpc scales \citep{Stockton87,Fu09a}. Nonetheless, four of our ENLR objects (SDSS\,J0126+1420, 0815+4304, 1051+6251, and 1253+2547) have extended \OIII\ luminosity greater than $5\times10^{41}$ erg s$^{-1}$ at radii greater than 10~kpc, satisfying the definition of EELRs. We consider an emission-line region extended if the offset between the \OIII\ centroid and the closest stellar nucleus is greater than $\sim$0.4\arcsec. It is interesting that ENLRs are {\em equally} represented in merging (13/26) and isolated (8/16) dpAGNs.

SDSS\,J1240+3534 was observed by OSIRIS, so the datacube has much better spatial resolution. Although SDSS\,J1240+3534 is a binary AGN (\S~\ref{sec:binary}), we include it here---in the ENLR category---because the ENLR produced the redshifted emission line and SNIFS would have resolved the two major spectral components ($\Delta\theta \sim$ 0.6\arcsec) through spectral decomposition. 

Six of the 21 ENLR sources have a secondary classification of ``ambiguous". These sources show resolved stellar components within 1.2\arcsec, and the seeing-limited spatially extended \OIII\ emission envelops the stellar components. Therefore, we classified them as ``ambiguous" because they could still be binary AGNs (Caveat \#2 at the beginning of the section).

\subsection{Unresolved Narrow-Line Regions: Aligned or Young Outflows?} \label{sec:unresolved}

\begin{figure*}[!t]
\plotone{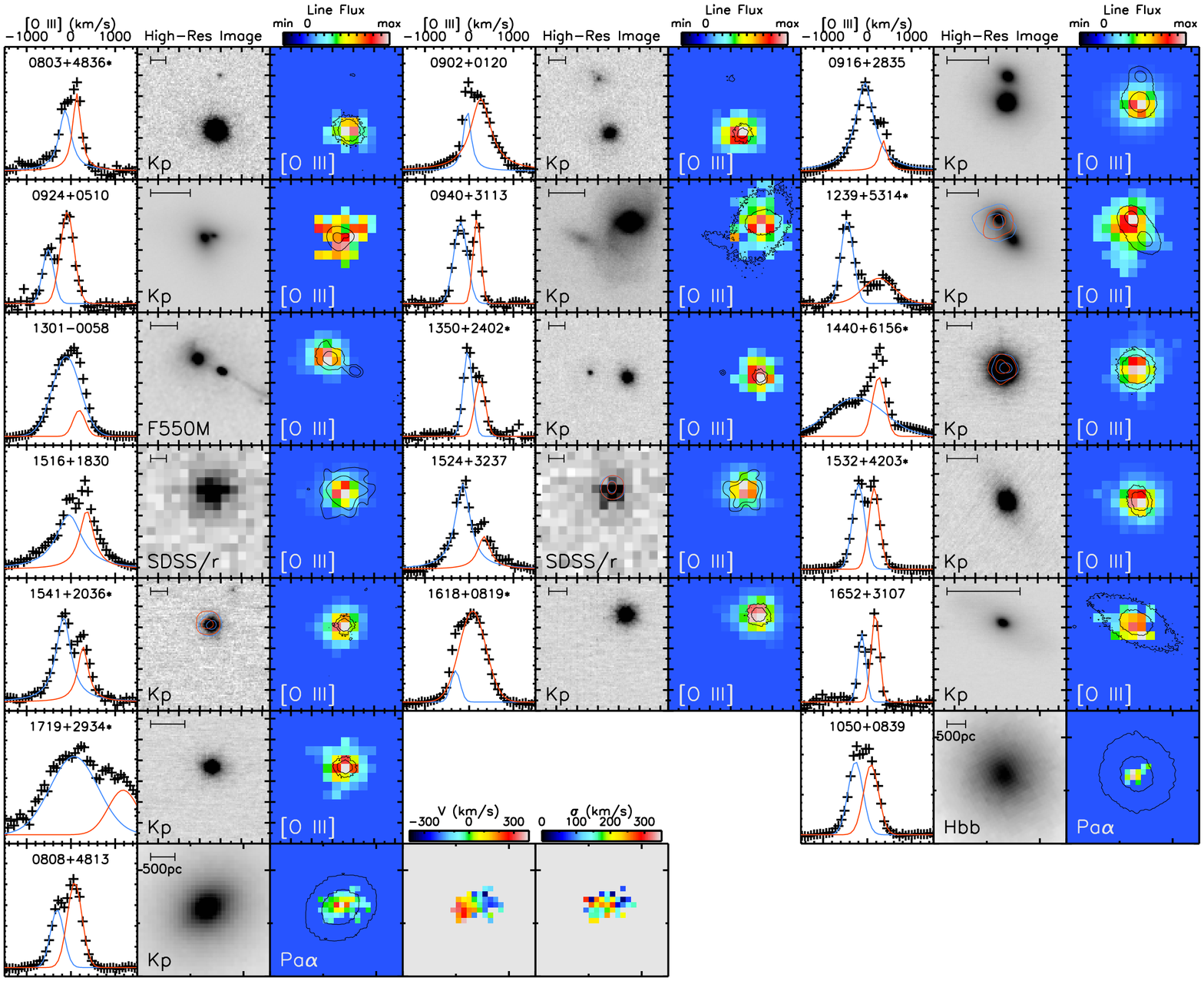}
\caption{Double-peaked AGNs with spatially unresolved kinematics. See Fig.~\ref{fig:binary} for SDSS\,J1151+4711. For each source, we show the \OIIIexpanded\ velocity profile from the SDSS, the best-available broad-band image, and the emission line flux map from the IFS datacube. N is up and E is left in all images. Like in Fig.~\ref{fig:binary}, for objects where SNIFS spectrally resolves the kinematic components, we overlay the emission-line fluxes from the blueshifted and redshifted components as blue and red contours in the broad-band images. Note that the $K_{\rm p}$ image of SDSS\,J1652+3107 does not show the companion clearly; see the contours in the associated line flux map for the position of the faint companion. The IFS data of SDSS\,J0808+4813 and SDSS\,J1050+0839 are from OSIRIS, so they are shown in last two rows. For SDSS\,J0808+4813, we also show the Pa$\alpha$ kinematic maps because it is resolved at 0.1\arcsec\ resolution. 
\label{fig:unresolved}} 
\end{figure*}

Finally we consider the category of unresolved narrow-line regions. For the 19 sources in this last category, the centroids of the two major spectral components of each source spatially overlap (Fig.~\ref{fig:unresolved}). It is possible that these sources can be spatially resolved by AO-aided integral-field spectroscopy with OSIRIS. But it is clear that a fraction of dpAGNs will remain unresolved even at 0.1\arcsec\ resolution, as seen in the example of SDSS\,J1050+0839. OSIRIS {\em did}, in fact, spatially resolve the gas kinematics in the nuclear NLR of SDSS\,J0808+4813. For consistency, we include SDSS\,J0808+4813 in the unresolved category because the spectral components are separated by less than $\sim$0.2\arcsec, so it would not have been resolved by SNIFS. Although the velocity gradient appears along the major axis of the host galaxy, SDSS\,J0808+4813 is unlikely a nuclear disk: because the $K$-band luminosity indicates a total stellar mass of $\sim2\times10^{10}$ \msun, the virial mass within the central 200~pc ($V^2 R/G$ = $1.8\times10^9$ ($V$/200~\kms)$^2$ ($R$/200~pc) M$_{\odot}$) seems too large to be gravitationally confined. We believe most of these spatially unresolved dpAGNs are aligned or young outflows.

Among the 11 merging galaxies in this category, SDSS\,J1151+4711 is a binary AGN (\S~\ref{sec:binary}), SDSS\,J0924+0510 is ambiguous ($\Delta\theta$ = 0.4\arcsec), SDSS\,J1239+5314 has an ENLR, and the rest appear to be single AGNs. Of course, we cannot exclude faint \OIII\ emission from the companions below our SNIFS sensitivity limit ($\simeq 1.4\pm1.0\times10^{-16}$\,erg\,s$^{-1}$\,cm$^{-2}$ \,arcsec$^{-2}$; \S~\ref{sec:snifs}) or AGNs without optical signatures.

\section{Integral-Field vs. Long-Slit Spectroscopy} \label{sec:comparison}

Six of the 42 IFS-observed sources overlap with the long-slit sample of \citet{Liu10b} and \citet{Shen11}. The overlapped sample includes 4 binary AGNs, 1 resolved NLR (similar to our ``ENLR" class), and 1 ambiguous source (similar to our ``unresolved" class) according to their classification. These sources are noted in Table~\ref{tab:ifu_sample} and are all in our ENLR category (Fig.~\ref{fig:enlr}). Our classifications differ from theirs for the same sources because of the complete spatial coverage of the IFS data. This result highlights the importance of integral-field spectroscopy in classifying dpAGNs. In the following, we discuss the overlapping sources in details.

We agree on the classification of SDSS\,J0400$-$0652 as an ENLR source. \citet{Shen11} showed two long-slit spectra at PA = $95^{\circ}$ and $173^{\circ}$; and only the spectrum at PA = $95^{\circ}$ shows the ENLR, in agreement with the gas morphology we see in the SNIFS datacube (Fig.~\ref{fig:enlr}). 

We classified SDSS\,J1108+0659 as an ambiguous ENLR source because the two nuclei, separated by 0.7\arcsec, are swamped by the much larger ($\sim$4\arcsec) ENLR. \citet{Liu10b} measure only a 0.9\arcsec\ separation between the \OIII\ components in their long-slit spectrum because the slit is aligned with the nuclei, and therefore, it missed the bulk of the emission from the ENLR.

Our IFS data suggest that the other three of the four supposed binary AGNs in their sample are likely ENLRs powered by single AGNs (SDSS\,J1146+5110, SDSS\,J1332+0606, and SDSS\,J1356+1026), although additional data are required to test both scenarios. For SDSS\,J1146+5110, the NE nucleus shows strong H$\alpha$ emission but non-detectable \OIII\ emission, so the gas is likely photoionized, at least partly, by star formation. The 2D spectrum of \citet{Liu10b} shows clearly three kinematic components in \OII, H$\alpha$, and \NII\ but only two prominent SW components in \OIII. The authors classified SDSS\,J1332+0606 and SDSS\,J1356+1026 as binary AGNs because the emission-line clouds in the ENLRs are aligned at similar PAs to the stellar components. Note that for both SDSS\,J1332+0606 and SDSS\,J1356+1026, the centroids of the \OIII\ emission are offset from the southern nuclei by a few kpcs and the emission seen at the locations of these nuclei are likely due to seeing smearing. Of course, we cannot rule out the possibility that the observed offset is due to blending between a weak nuclear NLR and a strong offset emission-line nebula, like in the case of SDSS\,J2333+0049 (Fig.~\ref{fig:enlr}). Future deep IFS data with higher spatial resolution can test this scenario.

\citet{Shen11} did not spatially resolve the ENLR of SDSS\,J2333+0049 because their long-slit PA of $117^{\circ}$ is almost perpendicular to the elongated orientation of the ENLR.

\section{Discussion} \label{sec:discuss}

\subsection{Statistics of Double-Peaked AGN Types} \label{sec:summary}

The majority of dpAGNs are due either to ENLRs or to small-scale gas kinematics. ENLRs account for half (21/42) of our IFS-observed dpAGNs in both merging (13/26) and isolated sources (8/16). We thus expect ENLRs around 50\% of dpAGNs. Unresolved gas kinematics account for 45\% (19/42) of the IFS-observed dpAGNs, which splits into 42\% (11/26) and 50\% (8/16) in merging and isolated AGNs, respectively. Taking into account the bias of the IFS sample, we expect 48\% of dpAGNs are due to unresolved nuclear gas kinematics.

We found four promising binary AGNs. Only two of these have a double-peaked line profile attributable to the relative velocity difference of the merging nuclei; for the other two the double-peaked profile is a result of gas kinematics. Binary AGNs thus account for 15\% (4/26) of the merging dpAGNs (that we have {\em resolved}). There are seven ambiguous cases for which we cannot rule out the binary AGN scenario, because of the limited spatial resolution of the SNIFS datacubes (Table~\ref{tab:ifu_sample}). If we assume that all of the ambiguous cases are binaries, then we get an upper limit of 42\% (11/26) for the binary AGN fraction in merging dpAGNs.

From our high-resolution imaging survey we found that only 29\% (31/106) of dpAGNs are mergers; and for the purpose of this discussion, let us assume that AGNs with a single stellar nucleus cannot be kpc-scale binary AGNs. Therefore, we conclude that the binary AGN fraction in dpAGNs is 4.5$-$12\%\footnote{The total percentage of the above three categories exceeds 100\% because 2$-$9 of the binary AGNs also belong to the other two categories.}. If we include the ambiguous cases, our result agrees with that of \citet{Shen11} ($\sim$10\%), although we have identified a different set of binary AGNs (see \S~\ref{sec:comparison}). In the next subsection, we discuss the binary AGN fraction in all SDSS AGNs.

\subsection{Binary AGN Fraction and Completeness of Double-Peaked Selection} \label{sec:bFrac}

From a Monte Carlo simulation assuming identical intrinsic relative velocities (500 \kms) and line widths (250 \kms) for all systems, \citet{Shen11} estimated that roughly 20\% of kpc-scale binary AGNs would have been selected as a double-peaked AGN from SDSS fiber spectra. In the following, we attempt to constrain the total binary fraction of all SDSS AGNs ($f_{\rm Bin}$) and the completeness of the double-peaked selection to kpc-scale binaries ($c_{\rm dp}$) using the four promising binary AGNs we have identified in a sample of 26 merging dpAGNs.

Because $\sim$1\% of AGNs are double-peaked \citep[$f_{\rm dp}$;][]{Wang09,Liu10a,Smith10} and $29^{+5}_{-4}$\% of dpAGNs are resolved mergers in our 0.1\arcsec-resolution images, the fraction of AGNs that are binaries where orbital motion has produced the line-splitting is $0.022_{-0.008}^{+0.025}$\% (2/26$\times$29\%$\times$1\%; errors have been propagated); and it should be the product of the total binary fraction and the double-peaked selection completeness:

\begin{equation}
f_{\rm dpBin} = f_{\rm Bin} c_{\rm dp}
\end{equation}

Similarly, the fraction of serendipitous binaries among AGNs where other mechanisms produced line splitting is $0.022_{-0.008}^{+0.025}$\% (2/24$\times$24/26$\times$29\%$\times$1\%). Assuming that binary AGN do not tend to drive physical mechanisms (such as outflows) that can produce double-peaked emission lines, the serendipitous binary fraction should be the product of the kinematic dpAGN\footnote{These are objects that are double-peaked because of gas kinematics instead of orbital motion.} fraction in all AGNs, the total binary fraction, and the double-peaked selection incompleteness ($1-c_{\rm dp}$):

\begin{equation}
f_{\rm seBin} \simeq f_{\rm dp} f_{\rm Bin} (1 - c_{\rm dp})
\end{equation}

\noindent
For simplicity, we have used the total dpAGN fraction ($f_{\rm dp}$) in place of non-binary dpAGN fraction, because 98\% of the dpAGNs are due to mechanisms other than binary orbital motion (\S~\ref{sec:summary}).

Combining the above two equations, we obtain:

\begin{equation}
f_{\rm Bin} = f_{\rm dpBin} + f_{\rm seBin}/f_{\rm dp} \simeq 100 f_{\rm seBin} = 2.2_{-0.8}^{+2.5}\% 
\end{equation}
\begin{equation}
c_{\rm dp} = f_{\rm dpBin}/f_{\rm Bin} \simeq f_{\rm dpBin}/(100f_{\rm seBin}) = 1.0_{-0.5}^{+1.6}\%
\end{equation}

Our nominal binary AGN fraction lies close to the upper limit of that of \citet{Shen11} (0.5$-$2.5\%), but the double-peaked selection completeness is an order-of-magnitude lower. This implies that the characteristic relative velocity of binary AGNs is lower than the 500 \kms\ assumed by \citet{Shen11}. 

There is a non-negligible probability that SMBHs in a merging pair are accreting simultaneously by chance, appearing as a binary AGN. If interactions do not enhance nuclear accretion, we would expect $\lesssim$0.3\% of AGNs are kpc-scale binaries. This ``chance" binary fraction is simply the product of the average AGN duty cycle \citep[$\lesssim$1\% for $M_{\rm BH} \gtrsim 3\times10^7$ \msun\ at $z = 0.5$;][]{Shankar09} and the $\sim$29\% merger fraction of AGNs (assuming it is the same as that of the dpAGNs). We adopted the 1\% duty cycle because the black hole masses of the type-1 dpAGNs are all greater than $3\times10^7$ \msun\ with a mean value of $3\times10^8$ \msun\ \citep{Shen11a}. The observed binary fraction ($f_{\rm Bin}$) is already five times higher than this ``chance" binary fraction even at the lower bound of the 1$\sigma$ confidence interval. This result provides evidence that galaxy interactions enhance AGN activities (i.e., increasing the AGN duty cycle from $\sim$1\% to $8_{-3}^{+8}$\%), although a larger sample of binary AGNs and a more sophisticated calculation incorporating AGN \OIII\ luminosity functions and merger rates \citep[e.g.,][]{Yu11} are needed to draw the final conclusion.

\subsection{The Formation of Extended Narrow-Line Regions} \label{sec:formation}

\begin{figure}
\plotone{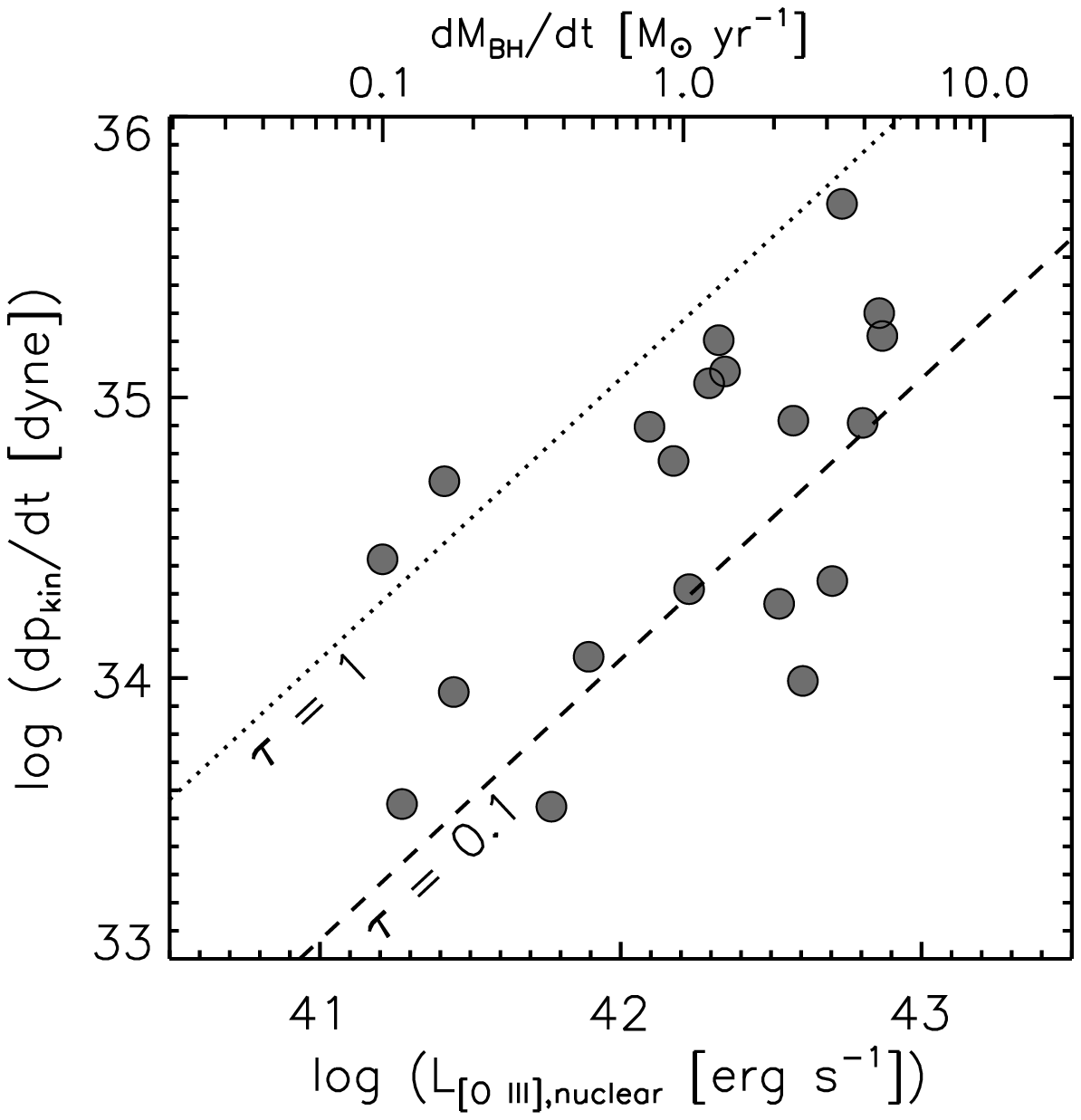}
\caption{
Momentum injection rate vs. nuclear \OIII\ luminosity (2\arcsec-diameter aperture) for the 20 SNIFS-identified ENLRs (Fig.~\ref{fig:enlr}). The top axis shows the corresponding black hole accretion rates, calculated assuming $L_{\rm bol} = 3500 L_{\rm [O~III]}$ \citep{Heckman04} and a radiative efficiency $\epsilon = 0.1$. The dotted and dashed lines indicate the momentum injection rates from radiation pressure of the current AGN luminosity at optical depths $\tau = 1.0$ and 0.1, respectively.
\label{fig:pdot_lo3}} 
\end{figure}

Although inefficient for identifying binary AGNs, the double-peaked selection technique is efficient for identifying ENLRs around AGNs (\S~\ref{sec:summary}): the SNIFS observations have accumulated 20 ENLRs at $z < 0.6$. Here we attempt to constrain the formation of such extended nebulae with the current data. 

The mass of a fully ionized cloud is proportional to the luminosity of hydrogen recombination lines at a given electron density ($n_e$): 

\begin{eqnarray}
M_{\rm H} & = & m_p n_e V f \nonumber \\
& = & m_p L_{\rm H\beta} / (\alpha_{\rm H\beta} n_e h\nu) \nonumber \\
& = & 6.8\times10^7~M_{\odot} (L_{\rm H\beta}/10^{40}~{\rm erg~s^{-1}}) (1~{\rm cm^{-3}}/n_e)
\end{eqnarray}

\noindent where $V$ is the total volume occupied by the cloud, $f$ is the filling factor, $m_p$ is the proton mass, $\alpha_{\rm H\beta}$ the case-B recombination coefficient of H$\beta$ \citep[$3.03\times10^{-14}$ cm$^3$~s$^{-1}$;][]{Osterbrock06}, and $h\nu$ the energy of an H$\beta$ photon ($4.09\times10^{-12}$ erg). Adopting an average \OIIIexpanded/\Hb\ ratio of 6 and $n_e = 1$ cm$^{-3}$, the total masses in the ENLRs range from $8\times10^7$ to $2\times10^{10}$ \msun. To avoid the ``classical" NLR in the inner part, we integrated only the pixels at distances greater than 1\arcsec\ from the nuclei. 

Further, we can estimate the total momentum ($p_{\rm kin} = M_H v$) of the ionized gas with the velocity maps. Finally, the size of the ENLR (\S~\ref{sec:size}) divided by the maximum line-of-sight velocity gives a dynamical timescale ($t = R_{NLR}/v_{max}$) between 8 and 70 Myrs, consistent with the typical AGN lifetime. Assuming the momentum has been accumulated throughout the dynamical timescale, we can estimate the rate of momentum injection ($\dot{p}_{\rm kin} \simeq p_{\rm kin}/t$), a diagnostic that can be compared with models.

It is possible that the radiation pressure from the AGN created the ENLRs, because the gas is believed to be dusty \citep{Dopita02, Groves04a}. Our sample is a mixture of type-1 and type-2 AGNs, so we used the nuclear \OIIIexpanded\ luminosity to infer the AGN bolometric luminosity \citep[$L_{\rm bol} = 3500 L_{\rm [O~III]}$;][]{Zakamska03, Heckman04}. The rate of momentum injection from radiation pressure is $\dot{p} = \tau L_{\rm bol}/c$, where $\tau$ is the effective optical depth and $c$ is the speed of light. Figure~\ref{fig:pdot_lo3} compares the measured rate of momentum injection with the expected rate of injection from radiation pressure at the current AGN luminosity. It shows that radiation pressure is sufficient to drive the outflow if the ENLR is moderately optically thick to the AGN radiation.

The uncertainties in this analysis are dominated by the lack of knowledge on the mass-weighted electron density. We have adopted a density of 1 cm$^{-3}$ because of the success of the two-phase photoionization model in fitting the emission lines of quasar EELRs \citep{Stockton02,Fu07a}. There, we have also shown that \OII\ or \SII\ doublets overestimate mass-weighted electron densities because high-density low-excitation regions dominate \OII\ and \SII\ emission. Because the momentum is inversely proportional to the density, the data points in Fig.~\ref{fig:pdot_lo3} easily can be offset by one dex along the vertical direction, compromising the conclusion. Detailed photoionization modeling on deeper spectra is needed to better constrain the formation of the ENLRs.

\subsection{Size$-$Luminosity Relation of Narrow-Line Regions} \label{sec:size}

\begin{figure}
\plotone{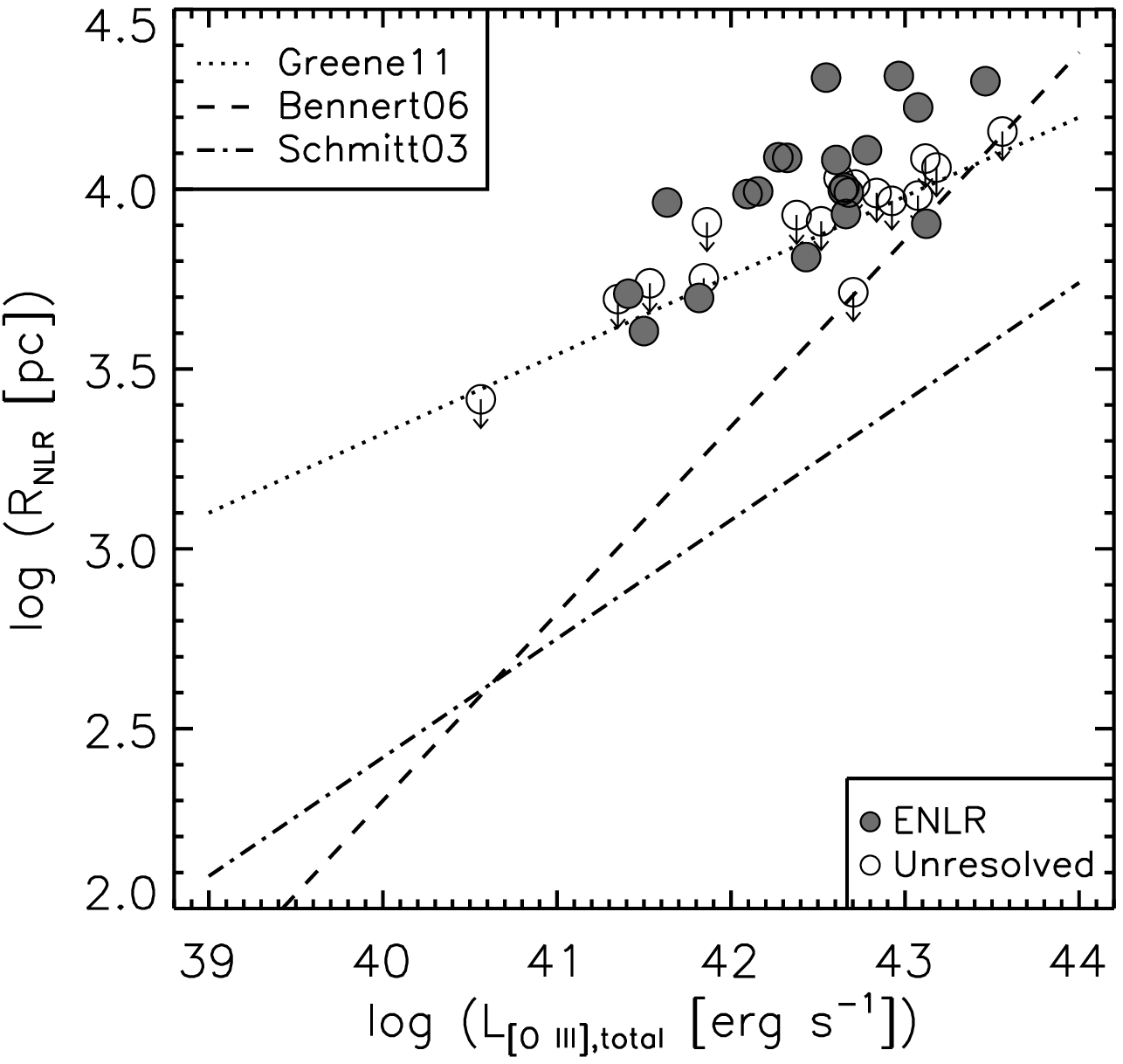}
\caption{Size-luminosity relation of dpAGN NLRs. The data points are SNIFS-observed dpAGNs. Spatially resolved NLRs (or ENLRs) are filled circles, while unresolved NLRs are open circles with downward arrows. Binary AGNs have been excluded. The dotted, dashed, and dot-dashed lines show the size-luminosity relations from \citet{Greene11}, \citet{Bennert02}, and \citet{Schmitt03}, respectively. 
\label{fig:size_lo3}} 
\end{figure}

AGN photoionized gas extending over kpc-scales has been detected around QSOs \citep{Stockton87} and Seyfert galaxies \citep{Unger87,Mulchaey96} for over three decades. Inspired by the size-luminosity relation of AGN broad-line regions \citep[e.g.,][]{Wandel99}, several more recent studies have suggested a similar correlation for the NLRs \citep[e.g.,][]{Bennert02,Schmitt03,Greene11}. Because of the importance of this relation to our understanding of the NLR, we re-examine this relation with the SNIFS data. Integral-field spectroscopy is well-suited for this kind of study because it combines the advantages of previous long-slit spectroscopy (better surface-brightness sensitivity) and \hst\ narrow-band imaging (more complete spatial coverage).

We determined the NLR radius as the longest distance from the nucleus to any lenslet with valid \OIIIexpanded\ detections, i.e., emission-line peak-to-noise ratios greater than 2.0. Figure~\ref{fig:size_lo3} plots the NLR radius as a function of \OIII\ luminosity integrated over the entire SNIFS field of view. We measured the sizes of both ENLRs and unresolved NLRs. For the latter, the measurements are considered as upper limits. 

We compare the best-fit relations from previous studies and our data. Because we reached a similar depth as the long-slit study of \citet{Greene11}, which is an order-of-magnitude deeper than those from \hst\ narrow-band imaging \citep[$> 2-12\times10^{-15}$ erg s$^{-1}$ cm$^{-2}$ arcsec$^{-2}$;][]{Bennert02}, our measurements are more consistent with the former than the latter. Considering the upper limits from the unresolved NLRs, the dispersion of this relation is at least 0.3 dex. 

\section{Summary}

We have used integral-field spectroscopy and 0.1\arcsec-resolution broad-band images to investigate the nature of double-peaked \OIII\ AGNs selected from the SDSS spectroscopic sample. Our main conclusions can be summarized as follows:

\begin{enumerate}

\item The relative velocity of a merging pair, each associated with an AGN NLR, accounts for only $\sim$2\% of the double-peaked \OIII\ AGNs. Gas kinematics produced $\sim$98\% of the double-peaked \OIII\ AGNs, which split almost evenly between spatially resolved and unresolved NLRs in seeing-limited data. We find the same fraction ($\sim$50\%) of extended narrow-line regions around mergers and isolated dpAGNs. 

\item We identified two binary AGNs from sources where gas kinematics produced the double-peaked \OIII\ lines. The relatively high fraction of such serendipitous binaries implies a low completeness ($\sim$1\%) of double-peaked selection to binary AGNs and a total binary AGN fraction of $2.2_{-0.8}^{+2.5}$\% in all SDSS AGNs at $z < 0.6$. The binary AGN fraction implies an AGN duty cycle ($8_{-3}^{+8}$\%) much higher than the average AGN duty cycle ($\lesssim$1\%), suggesting galaxy interactions enhance nuclear accretion.

\item About 29\% of the dpAGNs appear to be mergers at 0.1\arcsec-resolution. But multiple black holes are simultaneously accreting only in 15$-$42\% of these mergers. The upper limit included the seven ambiguous cases. Therefore, only 4$-$12\% of double-peaked SDSS AGNs could be binary AGNs.

\item The total mass of the ionized gas in the 20 extended narrow-line regions in the SNIFS sample range from $8\times10^7$ to $2\times10^{10}$ \msun, assuming an electron density of 1~cm$^{-3}$. We show that the radiation pressure from the AGN could have expelled enough gas to 10~kpc scales over the AGN lifetime and therefore created the extended narrow-line regions. 

\item We measure a size-luminosity relation for the NLRs consistent with that of \citet{Greene11} from ground-based long-slit spectra. There is at least 0.3~dex dispersion around the best-fit power-law.

\end{enumerate}

Although this work was first motivated by the belief that the double-peaked AGN sample offers the best candidates to search for kpc-scale binary AGNs, we and others \citep{Shen11,Rosario11} have shown that this selection is primarily identifying AGNs with secondary peaks due to ionized gas kinematics. Furthermore, through this study we have realized that the double-peaked sample offers very little advantage over more general AGN samples in terms of searching for binary AGNs, because the binary fraction in all AGNs ($\sim$2.4\%) is only slightly lower than that in dpAGNs ($\sim$4\%). Finally, we caution that because the binary AGNs that we found are either type-1$-$type-2 or type-2$-$type-2 pairs, we can not completely rule out the single AGN scenario: that the gas in one of the nuclei is photoionized by the AGN in the neighboring galaxy. The ultimate test awaits future high-resolution X-ray images from \chandra\ and/or radio images from the EVLA. 

\acknowledgments

It is a pleasure to thank Steven Rodney, Nick Moskovitz, Mark Willman, and Eric Gaidos for advice on SNIFS observations, and Jessica Lu, Tucker Jones, Peter Capak, and Nick Scoville for helpful discussions. We thank the referee for cogent comments that helped improve the paper.
ADM is a research fellow of the Alexander von Humboldt Foundation of Germany. 
AS was partially supported by NSF grant AST-0807900.
SGD was partially supported by NSF grant AST-0909182 and the Ajax Foundation.
GA was supported by the Director, Office of Science, Office of High Energy Physics, of the U.S. Department of Energy under Contract No. DE-AC02-05CH11231.
The authors wish to recognize and acknowledge the very significant cultural role and reverence that the summit of Mauna Kea has always had within the indigenous Hawaiian community. We are most fortunate to have the opportunity to conduct observations from this mountain.

{\it Facilities}: UH:2.2m (SNIFS), Keck:II (LGSAO/NIRC2, LGSAO/OSIRIS), Sloan 


\begin{thebibliography}
\expandafter\ifx\csname natexlab\endcsname\relax\def\natexlab#1{#1}\fi

\bibitem[{Aldering {et~al.}(2002)Aldering, Adam, Antilogus, Astier, Bacon,
  Bongard, Bonnaud, Copin, Hardin, Henault, Howell, Lemonnier, Levy, Loken,
  Nugent, Pain, Pecontal, Pecontal, Perlmutter, Quimby, Schahmaneche, Smadja,
  \& Wood-Vasey}]{Aldering02}
Aldering, G., {et~al.} 2002, in SPIE Conference Series, Vol. 4836, 61--72

\bibitem[{Aldering {et~al.}(2006)Aldering, Antilogus, Bailey, Baltay, Bauer,
  Blanc, Bongard, Copin, Gangler, Gilles, Kessler, Kocevski, Lee, Loken,
  Nugent, Pain, P{\'e}contal, Pereira, Perlmutter, Rabinowitz, Rigaudier,
  Scalzo, Smadja, Thomas, Wang, \& Weaver}]{Aldering06}
Aldering, G., {et~al.} 2006, \apj, 650, 510

\bibitem[{Bacon {et~al.}(1995)Bacon, Adam, Baranne, Courtes, Dubet, Dubois,
  Emsellem, Ferruit, Georgelin, Monnet, Pecontal, Rousset, \& Say}]{Bacon95}
Bacon, R., {et~al.} 1995, \aaps, 113, 347

\bibitem[{Baldwin {et~al.}(1981)Baldwin, Phillips, \& Terlevich}]{Baldwin81}
Baldwin, J.~A., Phillips, M.~M., \& Terlevich, R. 1981, \pasp, 93, 5

\bibitem[{Barnes \& Hernquist(1996)}]{Barnes96}
Barnes, J.~E., \& Hernquist, L. 1996, \apj, 471, 115

\bibitem[{Begelman {et~al.}(1980)Begelman, Blandford, \& Rees}]{Begelman80}
Begelman, M.~C., Blandford, R.~D., \& Rees, M.~J. 1980, Nature, 287, 307

\bibitem[{Bennert {et~al.}(2002)Bennert, Falcke, Schulz, Wilson, \&
  Wills}]{Bennert02}
Bennert, N., Falcke, H., Schulz, H., Wilson, A.~S., \& Wills, B.~J. 2002, \apj,
  574, L105

\bibitem[{Bianchi {et~al.}(2008)Bianchi, Chiaberge, Piconcelli, Guainazzi, \&
  Matt}]{Bianchi08}
Bianchi, S., Chiaberge, M., Piconcelli, E., Guainazzi, M., \& Matt, G. 2008,
  \mnras, 386, 105

\bibitem[{Boroson \& Lauer(2009)}]{Boroson09}
Boroson, T.~A., \& Lauer, T.~R. 2009, Nature, 458, 53

\bibitem[{Cameron(2011)}]{Cameron11}
Cameron, E. 2011, Publications of the Astronomical Society of Australia, 28,
  128

\bibitem[{Cardelli {et~al.}(1989)Cardelli, Clayton, \& Mathis}]{Cardelli89}
Cardelli, J.~A., Clayton, G.~C., \& Mathis, J.~S. 1989, \apj, 345, 245

\bibitem[{Chen {et~al.}(1989)Chen, Halpern, \& Filippenko}]{Chen89}
Chen, K., Halpern, J.~P., \& Filippenko, A.~V. 1989, \apj, 339, 742

\bibitem[{Comerford {et~al.}(2009)Comerford, Gerke, Newman, Davis, Yan, Cooper,
  Faber, Koo, Coil, Rosario, \& Dutton}]{Comerford09a}
Comerford, J.~M., {et~al.} 2009, \apj, 698, 956

\bibitem[{Davies(2007)}]{Davies07b}
Davies, R.~I. 2007, \mnras, 375, 1099

\bibitem[{Djorgovski(1991)}]{Djorgovski91}
Djorgovski, S. 1991, in The space distribution of quasars, Vol.~21, 349--353

\bibitem[{Djorgovski {et~al.}(1987)Djorgovski, Perley, Meylan, \&
  McCarthy}]{Djorgovski87}
Djorgovski, S., Perley, R., Meylan, G., \& McCarthy, P. 1987, \apj, 321, L17

\bibitem[{Dopita {et~al.}(2002)Dopita, Groves, Sutherland, Binette, \&
  Cecil}]{Dopita02}
Dopita, M.~A., Groves, B.~A., Sutherland, R.~S., Binette, L., \& Cecil, G.
  2002, \apj, 572, 753

\bibitem[{Dunkley {et~al.}(2009)Dunkley, Komatsu, Nolta, Spergel, Larson,
  Hinshaw, Page, Bennett, Gold, Jarosik, Weiland, Halpern, Hill, Kogut, Limon,
  Meyer, Tucker, Wollack, \& Wright}]{Dunkley09}
Dunkley, J., {et~al.} 2009, \apjs, 180, 306

\bibitem[{Eracleous {et~al.}(2011)Eracleous, Boroson, Halpern, \&
  Liu}]{Eracleous11}
Eracleous, M., Boroson, T.~A., Halpern, J.~P., \& Liu, J. 2011, preprint,
  arXiv:1106.2952

\bibitem[{Eracleous \& Halpern(2003)}]{Eracleous03}
Eracleous, M., \& Halpern, J.~P. 2003, \apj, 599, 886

\bibitem[{Eracleous {et~al.}(1997)Eracleous, Halpern, Gilbert, Newman, \&
  Filippenko}]{Eracleous97}
Eracleous, M., Halpern, J.~P., Gilbert, A.~M., Newman, J.~A., \& Filippenko,
  A.~V. 1997, \apj, 490, 216

\bibitem[{Fischer {et~al.}(2011)Fischer, Crenshaw, Kraemer, Schmitt, Mushotsky,
  \& Dunn}]{Fischer11}
Fischer, T.~C., Crenshaw, D.~M., Kraemer, S.~B., Schmitt, H.~R., Mushotsky,
  R.~F., \& Dunn, J.~P. 2011, \apj, 727, 71

\bibitem[{Fu {et~al.}(2011{\natexlab{a}})Fu, Myers, Djorgovski, \& Yan}]{Fu11a}
Fu, H., Myers, A.~D., Djorgovski, S.~G., \& Yan, L. 2011{\natexlab{a}}, \apj,
  733, 103

\bibitem[{Fu \& Stockton(2007)}]{Fu07a}
Fu, H., \& Stockton, A. 2007, \apj, 666, 794

\bibitem[{Fu \& Stockton(2009{\natexlab{a}})}]{Fu09a}
---. 2009{\natexlab{a}}, \apj, 690, 953

\bibitem[{Fu \& Stockton(2009{\natexlab{b}})}]{Fu09}
---. 2009{\natexlab{b}}, \apj, 696, 1693

\bibitem[{Fu {et~al.}(2011{\natexlab{b}})Fu, Zhang, Assef, Stockton, Myers,
  Yan, Djorgovski, Wrobel, \& Riechers}]{Fu11c}
Fu, H., {et~al.} 2011{\natexlab{b}}, \apjl, 740, L44 

\bibitem[{Gaskell(1983)}]{Gaskell83}
Gaskell, C.~M. 1983, in Liege International Astrophysical Colloquia, Vol.~24,
  473--477

\bibitem[{Gaskell(1996)}]{Gaskell96}
Gaskell, C.~M. 1996, \apj, 464, L107

\bibitem[{Gelderman \& Whittle(1994)}]{Gelderman94}
Gelderman, R., \& Whittle, M. 1994, \apjs, 91, 491

\bibitem[{Gerke {et~al.}(2007)Gerke, Newman, Lotz, Yan, Barmby, Coil,
  Conselice, Ivison, Lin, Koo, Nandra, Salim, Small, Weiner, Cooper, Davis,
  Faber, \& Guhathakurta}]{Gerke07}
Gerke, B.~F., {et~al.} 2007, \apj, 660, L23

\bibitem[{Green {et~al.}(2010)Green, Myers, Barkhouse, Mulchaey, Bennert, Cox,
  \& Aldcroft}]{Green10}
Green, P.~J., Myers, A.~D., Barkhouse, W.~A., Mulchaey, J.~S., Bennert, V.~N.,
  Cox, T.~J., \& Aldcroft, T.~L. 2010, \apj, 710, 1578

\bibitem[{Greene {et~al.}(2011)Greene, Zakamska, Ho, \& Barth}]{Greene11}
Greene, J.~E., Zakamska, N.~L., Ho, L.~C., \& Barth, A.~J. 2011, \apj, 732, 9

\bibitem[{Groves {et~al.}(2004)Groves, Dopita, \& Sutherland}]{Groves04a}
Groves, B.~A., Dopita, M.~A., \& Sutherland, R.~S. 2004, \apjs, 153, 9

\bibitem[{Heckman {et~al.}(2004)Heckman, Kauffmann, Brinchmann, Charlot,
  Tremonti, \& White}]{Heckman04}
Heckman, T.~M., Kauffmann, G., Brinchmann, J., Charlot, S., Tremonti, C., \&
  White, S. D.~M. 2004, \apj, 613, 109

\bibitem[{Hennawi {et~al.}(2006)Hennawi, Strauss, Oguri, Inada, Richards,
  Pindor, Schneider, Becker, Gregg, Hall, Johnston, Fan, Burles, Schlegel,
  Gunn, Lupton, Bahcall, Brunner, \& Brinkmann}]{Hennawi06}
Hennawi, J.~F., {et~al.} 2006, \aj, 131, 1

\bibitem[{Junkkarinen {et~al.}(2001)Junkkarinen, Shields, Beaver, Burbidge,
  Cohen, Hamann, \& Lyons}]{Junkkarinen01}
Junkkarinen, V., Shields, G.~A., Beaver, E.~A., Burbidge, E.~M., Cohen, R.~D.,
  Hamann, F., \& Lyons, R.~W. 2001, The Astrophysical Journal Letters, 549,
  L155

\bibitem[{Kauffmann {et~al.}(2003)Kauffmann, Heckman, Tremonti, Brinchmann,
  Charlot, White, Ridgway, Brinkmann, Fukugita, Hall, Ivezic, Richards, \&
  Schneider}]{Kauffmann03}
Kauffmann, G., {et~al.} 2003, \mnras, 346, 1055

\bibitem[{Kellermann {et~al.}(1989)Kellermann, Sramek, Schmidt, Shaffer, \&
  Green}]{Kellermann89}
Kellermann, K.~I., Sramek, R., Schmidt, M., Shaffer, D.~B., \& Green, R. 1989,
  \aj, 98, 1195

\bibitem[{Kewley {et~al.}(2006)Kewley, Groves, Kauffmann, \&
  Heckman}]{Kewley06}
Kewley, L.~J., Groves, B., Kauffmann, G., \& Heckman, T. 2006, \mnras, 372, 961

\bibitem[{Kochanek {et~al.}(1999)Kochanek, Falco, \& Mu{\~n}oz}]{Kochanek99}
Kochanek, C.~S., Falco, E.~E., \& Mu{\~n}oz, J.~A. 1999, \apj, 510, 590

\bibitem[{Komossa {et~al.}(2003)Komossa, Burwitz, Hasinger, Predehl, Kaastra,
  \& Ikebe}]{Komossa03}
Komossa, S., Burwitz, V., Hasinger, G., Predehl, P., Kaastra, J.~S., \& Ikebe,
  Y. 2003, \apj, 582, L15

\bibitem[{Komossa {et~al.}(2008)Komossa, Zhou, \& Lu}]{Komossa08}
Komossa, S., Zhou, H., \& Lu, H. 2008, \apj, 678, L81

\bibitem[{Kormendy \& Richstone(1995)}]{Kormendy95}
Kormendy, J., \& Richstone, D. 1995, \araa, 33, 581

\bibitem[{Koss {et~al.}(2011)Koss, Mushotzky, Treister, Veilleux, Vasudevan,
  Miller, Sanders, Schawinski, \& Trippe}]{Koss11}
Koss, M., {et~al.} 2011, \apj, 735, L42

\bibitem[{Krabbe {et~al.}(2004)Krabbe, Gasaway, Song, Iserlohe, Weiss, Larkin,
  Barczys, \& Lafreniere}]{Krabbe04}
Krabbe, A., Gasaway, T., Song, I., Iserlohe, C., Weiss, J., Larkin, J.~E.,
  Barczys, M., \& Lafreniere, D. 2004, in SPIE Conference Series, Vol. 5492,
  1403--1410

\bibitem[{Lantz {et~al.}(2004)Lantz, Aldering, Antilogus, Bonnaud, Capoani,
  Castera, Copin, Dubet, Gangler, Henault, Lemonnier, Pain, Pecontal, Pecontal,
  \& Smadja}]{Lantz04}
Lantz, B., {et~al.} 2004, in SPIE Conference Series, Vol. 5249, 146--155

\bibitem[{Larkin {et~al.}(2006)Larkin, Barczys, Krabbe, Adkins, Aliado, Amico,
  Brims, Campbell, Canfield, Gasaway, Honey, Iserlohe, Johnson, Kress,
  LaFreniere, Lyke, Magnone, Magnone, McElwain, Moon, Quirrenbach, Skulason,
  Song, Spencer, Weiss, \& Wright}]{Larkin06}
Larkin, J., {et~al.} 2006, in SPIE Conference Series, Vol. 6269, 42

\bibitem[{Liu {et~al.}(2010{\natexlab{a}})Liu, Greene, Shen, \&
  Strauss}]{Liu10b}
Liu, X., Greene, J.~E., Shen, Y., \& Strauss, M.~A. 2010{\natexlab{a}}, \apj,
  715, L30

\bibitem[{Liu {et~al.}(2010{\natexlab{b}})Liu, Shen, Strauss, \&
  Greene}]{Liu10a}
Liu, X., Shen, Y., Strauss, M.~A., \& Greene, J.~E. 2010{\natexlab{b}}, \apj,
  708, 427

\bibitem[{Liu {et~al.}(2011)Liu, Shen, Strauss, \& Hao}]{Liu11a}
Liu, X., Shen, Y., Strauss, M.~A., \& Hao, L. 2011, preprint, arXiv:1104.0950

\bibitem[{Markwardt(2009)}]{Markwardt09}
Markwardt, C.~B. 2009, in ASP Conference Series, Astronomical Data Analysis
  Software and Systems XVIII, D. A. Bohlender, D. Durand, and P. Dowler, eds.,
  Vol. 411, 251

\bibitem[{McGurk {et~al.}(2011)McGurk, Max, Rosario, Shields, Smith, \&
  Wright}]{McGurk11}
McGurk, R.~C., Max, C.~E., Rosario, D.~J., Shields, G.~A., Smith, K.~L., \&
  Wright, S.~A. 2011, \apjl, 738

\bibitem[{Mulchaey {et~al.}(1996)Mulchaey, Wilson, \& Tsvetanov}]{Mulchaey96}
Mulchaey, J.~S., Wilson, A.~S., \& Tsvetanov, Z. 1996, \apjs, 102, 309

\bibitem[{Myers {et~al.}(2007)Myers, Brunner, Richards, Nichol, Schneider, \&
  Bahcall}]{Myers07}
Myers, A.~D., Brunner, R.~J., Richards, G.~T., Nichol, R.~C., Schneider, D.~P.,
  \& Bahcall, N.~A. 2007, \apj, 658, 99

\bibitem[{Myers {et~al.}(2008)Myers, Richards, Brunner, Schneider, Strand,
  Hall, Blomquist, \& York}]{Myers08}
Myers, A.~D., Richards, G.~T., Brunner, R.~J., Schneider, D.~P., Strand, N.~E.,
  Hall, P.~B., Blomquist, J.~A., \& York, D.~G. 2008, \apj, 678, 635

\bibitem[{Osterbrock \& Ferland(2006)}]{Osterbrock06}
Osterbrock, D.~E., \& Ferland, G.~J. 2006, Astrophysics of gaseous nebulae and
  active galactic nuclei (Sausalito, CA: University Science Books, 2006)

\bibitem[{Richstone {et~al.}(1998)Richstone, Ajhar, Bender, Bower, Dressler,
  Faber, Filippenko, Gebhardt, Green, Ho, Kormendy, Lauer, Magorrian, \&
  Tremaine}]{Richstone98}
Richstone, D., {et~al.} 1998, Nature, 395, 14

\bibitem[{Rodriguez {et~al.}(2006)Rodriguez, Taylor, Zavala, Peck, Pollack, \&
  Romani}]{Rodriguez06}
Rodriguez, C., Taylor, G.~B., Zavala, R.~T., Peck, A.~B., Pollack, L.~K., \&
  Romani, R.~W. 2006, \apj, 646, 49

\bibitem[{Rosario {et~al.}(2011)Rosario, McGurk, Max, Shields, \&
  Smith}]{Rosario11}
Rosario, D.~J., McGurk, R.~C., Max, C.~E., Shields, G.~A., \& Smith, K.~L.
  2011, preprint, arXiv:1102.1733

\bibitem[{Rosario {et~al.}(2010)Rosario, Shields, Taylor, Salviander, \&
  Smith}]{Rosario10}
Rosario, D.~J., Shields, G.~A., Taylor, G.~B., Salviander, S., \& Smith, K.~L.
  2010, \apj, 716, 131

\bibitem[{Schlegel {et~al.}(1998)Schlegel, Finkbeiner, \& Davis}]{Schlegel98}
Schlegel, D.~J., Finkbeiner, D.~P., \& Davis, M. 1998, \apj, 500, 525

\bibitem[{Schmitt {et~al.}(2003)Schmitt, Donley, Antonucci, Hutchings, Kinney,
  \& Pringle}]{Schmitt03}
Schmitt, H.~R., Donley, J.~L., Antonucci, R. R.~J., Hutchings, J.~B., Kinney,
  A.~L., \& Pringle, J.~E. 2003, \apj, 597, 768

\bibitem[{Shankar {et~al.}(2009)Shankar, Weinberg, \&
  Miralda-Escud{\'e}}]{Shankar09}
Shankar, F., Weinberg, D.~H., \& Miralda-Escud{\'e}, J. 2009, \apj, 690, 20

\bibitem[{Shen {et~al.}(2011{\natexlab{a}})Shen, Liu, Greene, \&
  Strauss}]{Shen11}
Shen, Y., Liu, X., Greene, J.~E., \& Strauss, M.~A. 2011{\natexlab{a}}, \apj,
  735, 48

\bibitem[{Shen {et~al.}(2011{\natexlab{b}})Shen, Richards, Strauss, Hall,
  Schneider, Snedden, Bizyaev, Brewington, Malanushenko, Malanushenko, Oravetz,
  Pan, \& Simmons}]{Shen11a}
Shen, Y., {et~al.} 2011{\natexlab{b}}, \apjs, 194, 45

\bibitem[{Shields {et~al.}(2011)Shields, Rosario, Junkkarinen, Chapman,
  Bonning, \& Chiba}]{Shields11}
Shields, G.~A., Rosario, D.~J., Junkkarinen, V., Chapman, S.~C., Bonning,
  E.~W., \& Chiba, T. 2011, arXiv e-prints, 1109, 1524

\bibitem[{Smith {et~al.}(2010)Smith, Shields, Bonning, McMullen, Rosario, \&
  Salviander}]{Smith10}
Smith, K.~L., Shields, G.~A., Bonning, E.~W., McMullen, C.~C., Rosario, D.~J.,
  \& Salviander, S. 2010, \apj, 716, 866

\bibitem[{Stockton {et~al.}(2007)Stockton, Canalizo, Fu, \& Keel}]{Stockton07}
Stockton, A., Canalizo, G., Fu, H., \& Keel, W. 2007, \apj, 659, 195

\bibitem[{Stockton {et~al.}(2004)Stockton, Canalizo, Nelan, \&
  Ridgway}]{Stockton04}
Stockton, A., Canalizo, G., Nelan, E.~P., \& Ridgway, S.~E. 2004, \apj, 600,
  626

\bibitem[{Stockton \& MacKenty(1987)}]{Stockton87}
Stockton, A., \& MacKenty, J.~W. 1987, \apj, 316, 584

\bibitem[{Stockton {et~al.}(2002)Stockton, MacKenty, Hu, \& Kim}]{Stockton02}
Stockton, A., MacKenty, J.~W., Hu, E.~M., \& Kim, T.-S. 2002, \apj, 572, 735

\bibitem[{Toomre \& Toomre(1972)}]{Toomre72}
Toomre, A., \& Toomre, J. 1972, \apj, 178, 623

\bibitem[{Tsalmantza {et~al.}(2011)Tsalmantza, Decarli, Dotti, \&
  Hogg}]{Tsalmantza11}
Tsalmantza, P., Decarli, R., Dotti, M., \& Hogg, D.~W. 2011, arXiv e-prints,
  1106, 1180

\bibitem[{Unger {et~al.}(1987)Unger, Pedlar, Axon, Whittle, Meurs, \&
  Ward}]{Unger87}
Unger, S.~W., Pedlar, A., Axon, D.~J., Whittle, M., Meurs, E. J.~A., \& Ward,
  M.~J. 1987, \mnras, 228, 671

\bibitem[{Wandel {et~al.}(1999)Wandel, Peterson, \& Malkan}]{Wandel99}
Wandel, A., Peterson, B.~M., \& Malkan, M.~A. 1999, \apj, 526, 579

\bibitem[{Wang {et~al.}(2009)Wang, Chen, Hu, Mao, Zhang, \& Bian}]{Wang09}
Wang, J.-M., Chen, Y.-M., Hu, C., Mao, W.-M., Zhang, S., \& Bian, W.-H. 2009,
  \apj, 705, L76

\bibitem[{Wizinowich {et~al.}(2006)Wizinowich, Le~Mignant, Bouchez, Campbell,
  Chin, Contos, van Dam, Hartman, Johansson, Lafon, Lewis, Stomski, Summers,
  Brown, Danforth, Max, \& Pennington}]{Wizinowich06}
Wizinowich, P.~L., {et~al.} 2006, \pasp, 118, 297

\bibitem[{Xu \& Komossa(2009)}]{Xu09}
Xu, D., \& Komossa, S. 2009, \apj, 705, L20

\bibitem[{Yu {et~al.}(2011)Yu, Lu, Mohayaee, \& Colin}]{Yu11}
Yu, Q., Lu, Y., Mohayaee, R., \& Colin, J. 2011, arXiv e-prints, 1105, 1963

\bibitem[{Zakamska {et~al.}(2003)Zakamska, Strauss, Krolik, Collinge, Hall,
  Hao, Heckman, Ivezic, Richards, Schlegel, Schneider, Strateva, Vanden~Berk,
  Anderson, \& Brinkmann}]{Zakamska03}
Zakamska, N.~L., {et~al.} 2003, \aj, 126, 2125

\end{thebibliography}

\begin{deluxetable}{lcccccc} 
\tablewidth{0pt}
\tablecaption{Integral$-$Field Spectroscopy Sample 
\label{tab:ifulog}}
\tablehead{ 
\colhead{SDSS Name} & \colhead{IFS} & \colhead{Exptime} & \colhead{UT} & \colhead{Seeing} & \colhead{Depth} & \colhead{Image Source} \\
 & & (second) & & (\arcsec) & & \\
\colhead{(1)} & \colhead{(2)} & \colhead{(3)} & \colhead{(4)} & \colhead{(5)} & \colhead{(6)} & \colhead{(7)}
}
\startdata
012613.3$+$142013 &SNIFS/B\&R& 900&100809&0.7&0.9&       \nd\\
040001.6$-$065254 &   SNIFS/R&1800&100809&0.7&3.2&OSIRIS/Hbb\\
080315.7$+$483603*&SNIFS/B\&R& 900&110326&0.4&1.6&  NIRC2/Kp\\
080841.2$+$481352 &OSIRIS/Kn2&1800&100307&0.7&\nd&  NIRC2/Kp\\
081507.4$+$430427*&SNIFS/B\&R&1200&110329&0.7&3.8&  NIRC2/Kp\\
090246.9$+$012028 &SNIFS/B\&R&1200&110329&0.7&1.8&  NIRC2/Kp\\
091646.0$+$283527 &SNIFS/B\&R&1200&110329&0.6&4.1&  NIRC2/Kp\\
092455.2$+$051052 &SNIFS/B\&R&3600&110329&0.7&2.8&  NIRC2/Kp\\
094032.3$+$311329 &SNIFS/B\&R&1800&110329&1.1&1.8&  NIRC2/Kp\\
095207.6$+$255257*&SNIFS/B\&R&2400&110326&0.4&0.6&OSIRIS/Hbb\\
105052.5$+$083935 &OSIRIS/Kn3&2400&100307&0.7&\nd&  NIRC2/Kp\\
105104.5$+$625159*&SNIFS/B\&R&1200&110326&0.5&1.0&  NIRC2/Kp\\
110851.0$+$065901 &SNIFS/B\&R&1200&110329&0.8&2.1&  NIRC2/Kp\\
114642.5$+$511030 &SNIFS/B\&R&1200&110326&0.4&1.5&  NIRC2/Kp\\
115106.7$+$471158*&SNIFS/B\&R&1200&110326&0.6&0.6&  NIRC2/Kp\\
115523.7$+$150757*&SNIFS/B\&R& 900&101121&0.5&3.6&  NIRC2/Kp\\
123915.4$+$531415*&SNIFS/B\&R&1800&100804&0.5&1.2&  NIRC2/Kp\\
124037.8$+$353437 &OSIRIS/Kn3&1800&100307&0.7&\nd&OSIRIS/Hbb\\
124859.7$-$025731*&SNIFS/B\&R&2400&110326&0.4&0.6&  NIRC2/Kp\\
125327.5$+$254747*&SNIFS/B\&R&3600&110329&0.7&0.6&  NIRC2/Kp\\
130128.8$-$005804 &SNIFS/B\&R& 600&100808&0.6&6.3& ACS/F550M\\
130724.1$+$460401*&SNIFS/B\&R&1800&100808&0.5&0.5&  NIRC2/Kp\\
133226.3$+$060627 &SNIFS/B\&R&2400&110326&0.5&3.4&  NIRC2/Kp\\
135024.7$+$240251*&SNIFS/B\&R&2400&110326&0.8&1.1&  NIRC2/Kp\\
135646.1$+$102609 &SNIFS/B\&R& 900&100804&0.7&8.2&  NIRC2/Kp\\
140816.0$+$015528 &SNIFS/B\&R&3600&100809&0.5&1.2&       \nd\\
144012.8$+$615633*&SNIFS/B\&R&1200&110329&0.6&0.9&  NIRC2/Kp\\
150243.1$+$111557*&SNIFS/B\&R&1800&100804&0.8&1.3&  NIRC2/Kp\\
150437.7$+$541150*&SNIFS/B\&R&1200&110329&0.6&1.5&  NIRC2/Kp\\
151656.6$+$183022 &SNIFS/B\&R& 900&100809&0.6&0.9&       \nd\\
151735.2$+$214533*&SNIFS/B\&R&3600&100804&0.7&0.5&  NIRC2/Kp\\
152431.4$+$323751 &SNIFS/B\&R&1800&100809&0.6&0.3&       \nd\\
153231.8$+$420343*&SNIFS/B\&R&1200&110329&0.7&1.0&  NIRC2/Kp\\
154107.8$+$203609*&SNIFS/B\&R&3600&100806&0.6&1.0&  NIRC2/Kp\\
161826.9$+$081951*&SNIFS/B\&R& 900&100809&0.5&1.6&  NIRC2/Kp\\
161847.9$+$215925*&SNIFS/B\&R&1500&100809&0.6&0.8&  NIRC2/Kp\\
164713.4$+$383140 &SNIFS/B\&R&1800&100809&0.6&1.0&       \nd\\
165206.1$+$310708 &SNIFS/B\&R&1800&100806&0.6&2.3&  NIRC2/Kp\\
171930.6$+$293413*&SNIFS/B\&R&1200&110329&0.7&1.5&  NIRC2/Kp\\
172049.2$+$310646 &SNIFS/B\&R&1500&100804&0.5&1.6&  NIRC2/Kp\\
233313.2$+$004912 &SNIFS/B\&R& 900&100808&0.5&2.2&  NIRC2/Kp\\
235256.6$+$001155 &SNIFS/B\&R& 900&100809&0.6&1.4&  NIRC2/Kp\\
\enddata
\tablecomments{
Column 1: J2000 designation. type-1s are indicated by stars.
Column 2: Name of the Integral-Field Spectrograph (IFS).
Column 3: Total exposure time of the spectrum.
Column 4: UT date (yymmdd) of the spectrum.
Column 5: Outside seeing at 5000\AA\ during the spectroscopic observation
(http://kiloaoloa.soest.hawaii.edu/current/seeing/). 
Column 6: \OIII\ surface brightness depth of the reduced SNIFS datacube in 10$^{-16}$ erg~s$^{-1}$~cm$^{-2}$~arcsec$^{-2}$, see \S~\ref{sec:snifs}.
Column 7: Source of the high-resolution broad-band image.
}
\end{deluxetable}

\begin{deluxetable}{lccccccccccc} 
\rotate
\tabletypesize{\footnotesize}
\tablewidth{0pt}
\tablecaption{Properties of Double$-$peaked [O\,{\sc iii}] AGNs with Integral-Field
Spectroscopy
\label{tab:ifu_sample}}
\tablehead{ 
\colhead{SDSS Name} & \colhead{$z$} & \colhead{$\Delta V$} & 
\colhead{$L_{\rm [O III]}^b$} & \colhead{$L_{\rm [O III]}^r$} & \colhead{FWHM$^b$} &
\colhead{FWHM$^r$} & \colhead{$R$} & \colhead{$\Delta\theta$} &
\colhead{$\Delta S$} & \colhead{Secondary Class} & \colhead{S11 Class} \\
 & & (\kms) & log(\lsun) & log(\lsun) & (\kms) & (\kms) & & (\arcsec) &
 (kpc) & \\
\colhead{(1)} & \colhead{(2)} & \colhead{(3)} & \colhead{(4)} & 
\colhead{(5)} & \colhead{(6)} & \colhead{(7)} & \colhead{(8)} &
\colhead{(9)} & \colhead{(10)} & \colhead{(11)} & \colhead{(12)}
}
\startdata
\multicolumn{12}{c}{Orbital Motion} \nl
095207.6$+$255257*&0.339& 442&8.55&8.41& 682& 249&  \nd&1.00& 4.8& \nd      & \nd       \\
150243.1$+$111557*&0.390& 657&9.84&8.96& 384& 354&   36&1.39& 7.4& ENLR     & \nd       \\
\multicolumn{12}{c}{Extended Narrow-Line Region} \nl
012613.3$+$142013 &0.573& 580&8.72&9.35& 426& 334& 4698& \nd& \nd& \nd      & \nd       \\
040001.6$-$065254 &0.171& 391&8.43&8.73& 344& 397&   10& \nd& \nd& \nd      & NLR       \\
081507.4$+$430427*&0.510& 536&8.96&9.52& 419& 509&  106&1.95&12.0& \nd      & \nd       \\
105104.5$+$625159*&0.436& 364&9.25&8.19& 847& 167&    1&1.52& 8.6& \nd      & \nd       \\
110851.0$+$065901 &0.182& 204&8.74&8.27& 686& 159&   28&0.73& 2.2& ambiguous& binary    \\
114642.5$+$511030 &0.130& 287&8.03&8.17& 183& 264&  \nd&2.84& 6.6& \nd      & binary    \\
115523.7$+$150757*&0.287& 480&8.91&8.50& 928& 265&    2&0.58& 2.5& ambiguous& \nd       \\
124037.8$+$353437 &0.161& 491&8.79&8.20& 899& 338&   28&0.19& 0.5& binary   & \nd       \\
124859.7$-$025731*&0.487& 380&8.58&9.04&  92& 362&  \nd&0.53& 3.2& ambiguous& \nd       \\
125327.5$+$254747*&0.483& 688&8.74&8.25&1130& 148&   32&0.72& 4.3& ambiguous& \nd       \\
130724.1$+$460401*&0.352& 518&8.30&8.32& 629& 208&  \nd&2.37&11.7& \nd      & \nd       \\
133226.3$+$060627 &0.207& 409&7.79&8.04& 271& 439&  \nd&1.58& 5.4& \nd      & binary    \\
135646.1$+$102609 &0.123& 383&8.85&9.19& 397& 411&  128&1.32& 2.9& \nd      & binary    \\
140816.0$+$015528 &0.166& 339&7.64&7.67& 167& 216&   11& \nd& \nd& \nd      & \nd       \\
150437.7$+$541150*&0.305& 719&9.02&8.21& 362& 351&  \nd& \nd& \nd& \nd      & \nd       \\
151735.2$+$214533*&0.399& 324&8.62&8.06& 781& 124&  \nd&1.10& 5.9& ambiguous& \nd       \\
161847.9$+$215925*&0.334& 365&8.84&8.49& 328& 140&    5& \nd& \nd& \nd      & \nd       \\
164713.4$+$383140 &0.164& 340&7.52&7.49& 181& 185&  \nd& \nd& \nd& \nd      & \nd       \\
172049.2$+$310646 &0.095& 389&7.69&7.29& 231& 416&  \nd&0.38& 0.7& ambiguous& \nd       \\
233313.2$+$004912 &0.170& 434&8.29&8.67& 370& 449& 1234& \nd& \nd& \nd      & ambiguous \\
235256.6$+$001155 &0.167& 322&7.98&7.86& 287& 212&  \nd& \nd& \nd& \nd      & \nd       \\
\multicolumn{12}{c}{Unresolved Narrow-Line Region} \nl
080315.7$+$483603*&0.635& 262&8.98&8.92& 285& 157&    7&2.63&18.0& \nd      & \nd       \\
080841.2$+$481352 &0.124& 396&7.62&7.82& 320& 359&  \nd& \nd& \nd& \nd      & \nd       \\
090246.9$+$012028 &0.513& 304&8.87&9.37& 132& 526&   34&2.71&16.8& \nd      & \nd       \\
091646.0$+$283527 &0.142& 418&9.05&8.18& 473& 120&   13&1.23& 3.1& \nd      & \nd       \\
092455.2$+$051052 &0.150& 429&7.25&7.57& 261& 333&  \nd&0.44& 1.2& ambiguous& \nd       \\
094032.3$+$311329 &0.170& 374&7.71&7.49& 322& 132&  \nd&2.45& 7.1& \nd      & \nd       \\
105052.5$+$083935 &0.169& 351&8.19&8.18& 366& 369&  \nd& \nd& \nd& \nd      & \nd       \\
115106.7$+$471158*&0.318& 644&9.20&9.33& 774& 667&    3&1.29& 6.0& binary   & \nd       \\
123915.4$+$531415*&0.202& 715&7.87&7.76& 297& 825&   75&1.26& 4.2& ENLR     & \nd       \\
130128.8$-$005804 &0.246& 292&9.11&8.26& 731& 283&   17&1.40& 5.4& \nd      & \nd       \\
135024.7$+$240251*&0.557& 277&8.69&8.57& 190& 227&  \nd&1.80&11.6& \nd      & \nd       \\
144012.8$+$615633*&0.275& 505&8.89&8.42&1494& 298&    2& \nd& \nd& ENLR     & \nd       \\
151656.6$+$183022 &0.580& 400&9.77&9.53& 693& 354&26314& \nd& \nd& \nd      & \nd       \\
152431.4$+$323751 &0.629& 480&9.41&8.92& 378& 307&  \nd& \nd& \nd& \nd      & \nd       \\
153231.8$+$420343*&0.209& 353&8.51&8.44& 290& 251&  \nd& \nd& \nd& \nd      & \nd       \\
154107.8$+$203609*&0.508& 441&8.96&8.63& 374& 241&  \nd&2.00&12.3& \nd      & \nd       \\
161826.9$+$081951*&0.446& 384&8.62&9.54& 195& 689&  137& \nd& \nd& \nd      & \nd       \\
165206.1$+$310708 &0.075& 307&6.59&6.73& 137& 155&  \nd&2.96& 4.2& \nd      & \nd       \\
171930.6$+$293413*&0.180&1105&8.12&7.67&1265& 776&    2& \nd& \nd& \nd      & \nd       \\
\enddata
\tablecomments{
Objects are grouped according to the origin of the double-peaked \OIII\ lines.
Column 1: J2000 designation. type-1s are indicated by stars.
Column 2: Redshift.
Column 3: Velocity splitting between \OIII\,$\lambda 5007$
components: $\Delta V/c = [(1+z_r)^2/(1+z_b)^2-1]/[(1+z_r)^2/(1+z_b)^2+1]$, where $z_r$ and $z_b$ are the redshifts of the redshifted and blueshifted \OIII\ components, respectively.
Columns 4 and 5: \OIIIexpanded\ luminosity in log(\lsun) for the
blueshifted ($b$) and redshifted ($r$) line, corrected for Galactic extinction.
Columns 6 and 7: \OIIIexpanded\ FWHMs, corrected for the $\sigma =
65$ \kms\ instrumental broadening.
Column 8: \citet{Kellermann89} radio loudness, $R = F_{\nu,\rm 5GHz}/F_{\nu,\rm 4400A}$.
Columns 9 and 10: Projected angular separation (arcsec) and physical
separation (kpc) between the main components in a merging system.
Column 11: Secondary classification (\S~\ref{sec:class}).
Column 12: Classification from \citet{Shen11}.
}
\end{deluxetable}

\clearpage

\appendix

\section{High-Resolution Images of Double-Peaked AGNs}

\begin{figure*}[!tb]
\plotone{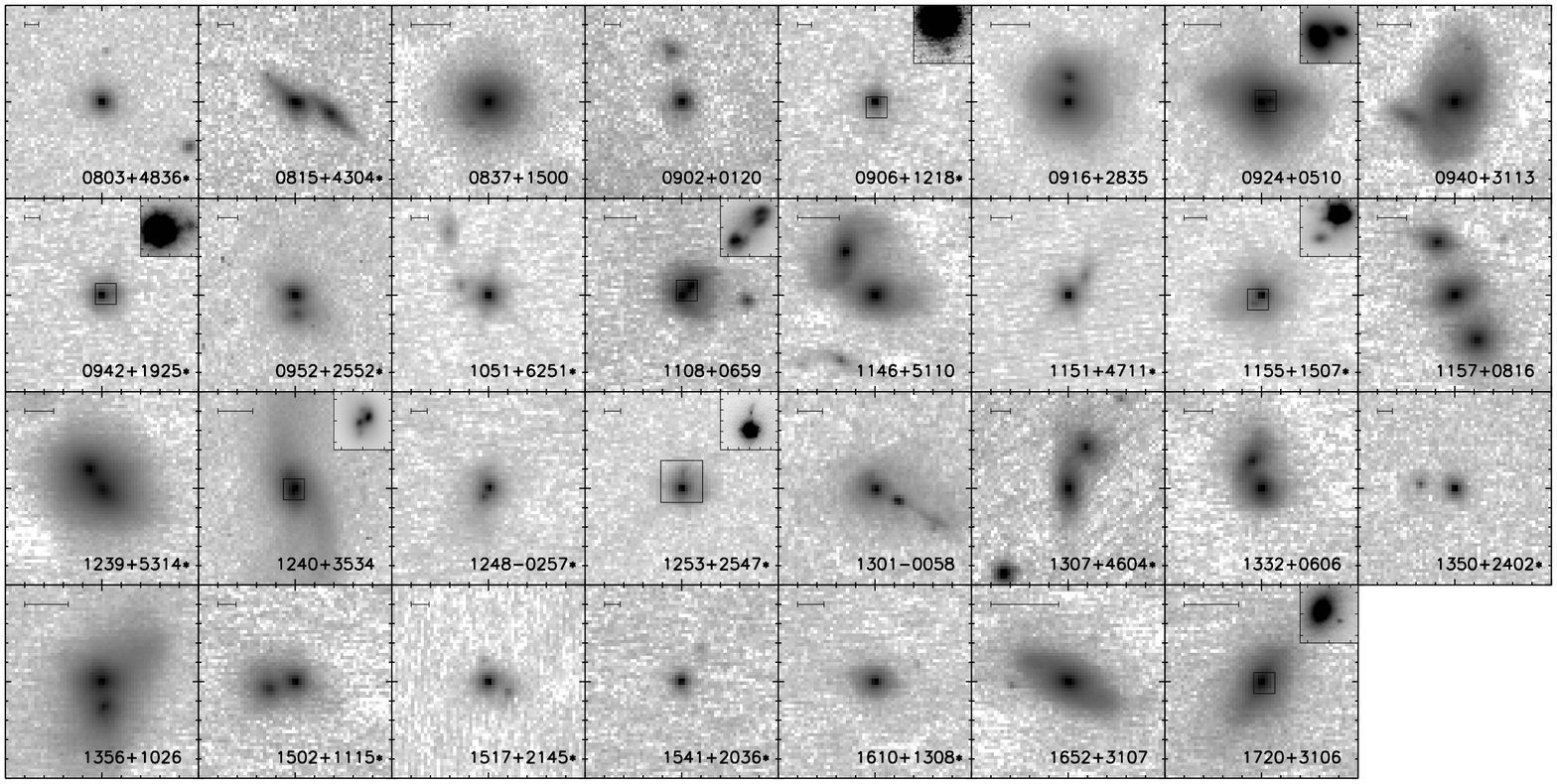}
\caption{Keck\,II/LGSAO or \hst\ images of dpAGNs that have companions within 3\arcsec. Images are displayed in $asinh$ (the inverse hyperbolic sine function) scales to bring up low surface brightness features. North is up and east is to the left (10\arcsec\ stamps with 1\arcsec\ tickmarks). The scale bar indicates a transverse separation of 5~kpc. The insets show the components in linear scales for compact systems ($\Delta\theta <$ 1\arcsec). The open boxes in the main panel delineate the regions covered by the insets. The tickmarks in the insets are spaced in 0.4\arcsec. 
\label{fig:imglog_merger}} 
\end{figure*}

\begin{figure*}[!t]
\plotone{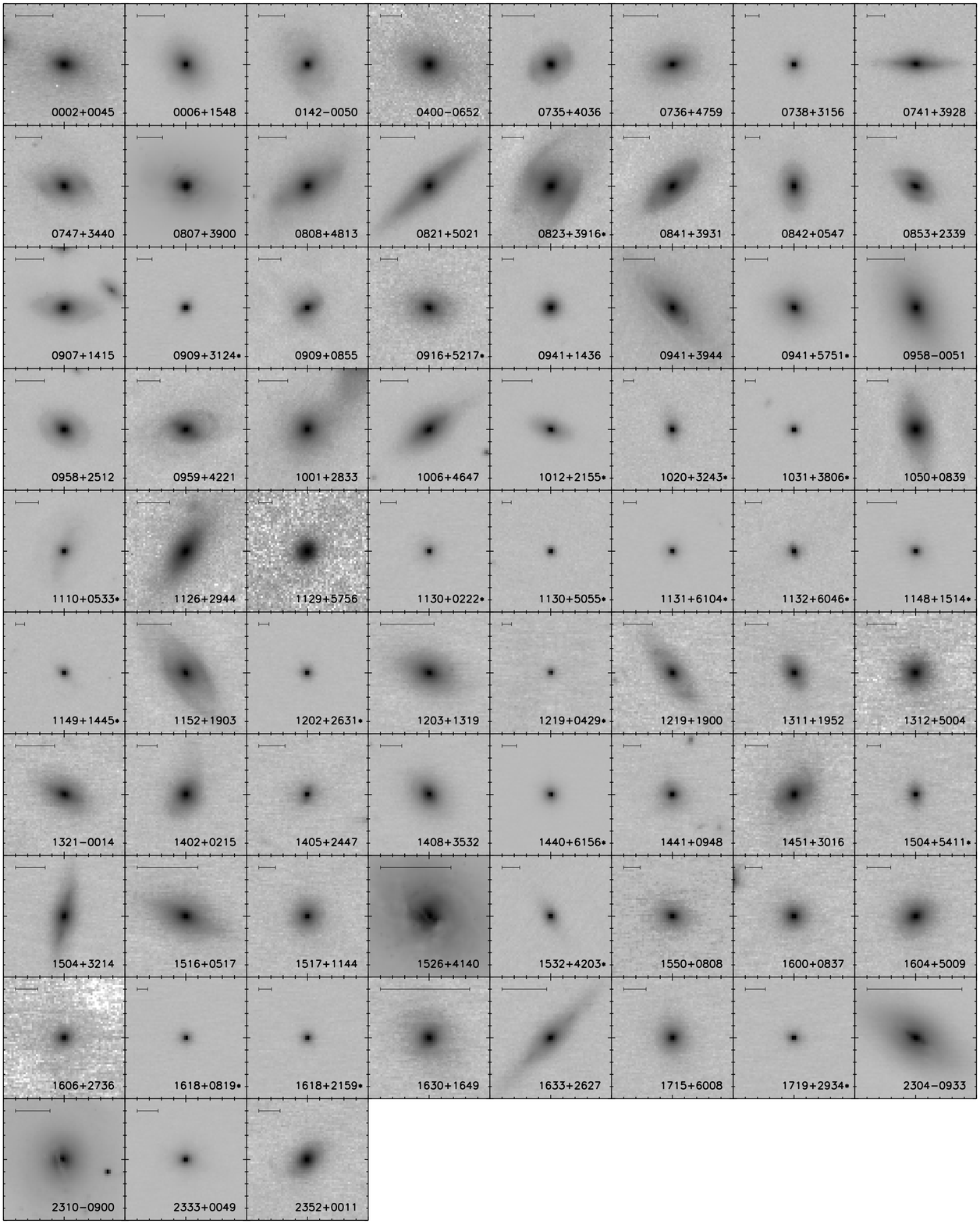}
\caption{Same as Fig.~\ref{fig:imglog_merger}, but for dpAGNs that do not have companions within 3\arcsec. The scale bar indicates a transverse separation of 5~kpc except for SDSS\,J0807+3900 and SDSS\,J1526+4140, where it indicates 1~kpc.
\label{fig:imglog_single}} 
\end{figure*}

We show the high-resolution images of the 106 dpAGNs from our Keck LGSAO program and the \hst\ archive in Figures~\ref{fig:imglog_merger} \& \ref{fig:imglog_single} for objects with and without companions within 3\arcsec, respectively. Tables~\ref{tab:imglog_merger} \& \ref{tab:imglog_single} list the properties and observing details.

\clearpage

\begin{deluxetable}{lccccccccccc} 
\rotate
\tablewidth{0pt}
\tablecaption{Double-Peaked \OIII\ AGNs with Companions within 3\arcsec
\label{tab:imglog_merger}}
\tablehead{ 
\colhead{SDSS Name} & \colhead{$z$} & \colhead{$\Delta$$V$} & \colhead{$R$} &
\colhead{$\Delta\theta$} & \colhead{$\Delta S$} &
\colhead{Imager} & \colhead{Exptime} & \colhead{UT} & \colhead{Seeing} &
\colhead{IFS} & \colhead{Ref}\\
 & & (\kms) & &
 (\arcsec) & (kpc) &
 & (second) & & (\arcsec) & 
 & \\
\colhead{(1)} & \colhead{(2)} & \colhead{(3)} & \colhead{(4)} & 
\colhead{(5)} & \colhead{(6)} & \colhead{(7)} & \colhead{(8)} &
\colhead{(9)} & \colhead{(10)} & \colhead{(11)} & \colhead{(12)} 
}
\startdata
080315.7$+$483603*&0.635& 262&    7&2.63&18.0&   NIRC2/Kp& 420&110105&0.6&1&3    \\
081507.4$+$430427*&0.510& 536&  106&1.95&12.0&   NIRC2/Kp& 540&110105&0.6&1&3    \\
083713.5$+$150037 &0.141& 387&    7&1.37& 3.4&   NIRC2/Kp& 480&110106&0.9&0&1    \\
090246.9$+$012028 &0.513& 304&   34&2.71&16.8&   NIRC2/Kp& 600&110106&0.8&1&3    \\
090615.9$+$121846*&0.644& 357&  \nd&0.61& 4.2&   NIRC2/Kp& 448&110105&0.5&0&3    \\
091646.0$+$283527 &0.142& 418&   13&1.23& 3.1&   NIRC2/Kp& 660&110106&1.0&1&1    \\
092455.2$+$051052 &0.150& 429&  \nd&0.44& 1.2&   NIRC2/Kp& 660&110106&1.0&1&3    \\
094032.3$+$311329 &0.170& 374&  \nd&2.45& 7.1&   NIRC2/Kp& 840&110106&1.0&1&1    \\
094236.7$+$192541*&0.540& 515&    1&0.44& 2.8&   NIRC2/Kp& 640&110105&0.5&0&3    \\
095207.6$+$255257*&0.339& 442&  \nd&1.00& 4.8& OSIRIS/Hbb& 480&100306&0.7&1&3    \\
105104.5$+$625159*&0.436& 364&    1&1.52& 8.6&   NIRC2/Kp& 600&110105&0.5&1&3    \\
110851.0$+$065901 &0.182& 204&   28&0.73& 2.2&   NIRC2/Kp& 192&100604&0.3&1&1    \\
114642.5$+$511030 &0.130& 287&  \nd&2.84& 6.6&   NIRC2/Kp& 704&100604&0.4&1&1    \\
115106.7$+$471158*&0.318& 644&    3&1.29& 6.0&   NIRC2/Kp& 600&110105&0.4&1&3    \\
115523.7$+$150757*&0.287& 480&    2&0.58& 2.5&   NIRC2/Kp& 352&100603&0.7&1&3    \\
115715.0$+$081632 &0.201& 375&  \nd&2.61& 8.6&   NIRC2/Kp& 448&100603&0.7&0&3    \\
123915.4$+$531415*&0.202& 715&   75&1.26& 4.2&   NIRC2/Kp&1088&100604&0.3&1&3    \\
124037.8$+$353437 &0.161& 491&   28&0.19& 0.5& OSIRIS/Hbb& 960&100307&0.7&1&1    \\
124859.7$-$025731*&0.487& 380&  \nd&0.53& 3.2&   NIRC2/Kp& 768&100604&0.4&1&3    \\
125327.5$+$254747*&0.483& 688&   32&0.72& 4.3&   NIRC2/Kp& 420&110105&0.4&1&3    \\
130128.8$-$005804 &0.246& 292&   17&1.40& 5.4&  ACS/F550M&2148&030809&\nd&1&1    \\
130724.1$+$460401*&0.352& 518&  \nd&2.37&11.7&   NIRC2/Kp& 448&100604&0.3&1&3    \\
133226.3$+$060627 &0.207& 409&  \nd&1.58& 5.4&   NIRC2/Kp& 768&100604&0.4&1&1,3  \\
135024.7$+$240251*&0.557& 277&  \nd&1.80&11.6&   NIRC2/Kp& 420&110105&0.4&1&3    \\
135646.1$+$102609 &0.123& 383&  128&1.32& 2.9&   NIRC2/Kp& 704&100603&0.8&1&1    \\
150243.1$+$111557*&0.390& 657&   36&1.39& 7.4&   NIRC2/Kp& 448&100603&0.7&1&3    \\
151735.2$+$214533*&0.399& 324&  \nd&1.10& 5.9&   NIRC2/Kp& 960&100604&0.4&1&3    \\
154107.8$+$203609*&0.508& 441&  \nd&2.00&12.3&   NIRC2/Kp& 448&100604&0.3&1&3    \\
161027.4$+$130807*&0.229& 326&  \nd&2.35& 8.6&   NIRC2/Kp& 192&100604&0.3&0&3    \\
165206.1$+$310708 &0.075& 307&  \nd&2.96& 4.2&   NIRC2/Kp& 320&100603&0.6&1&1,2  \\
172049.2$+$310646 &0.095& 389&  \nd&0.38& 0.7&   NIRC2/Kp& 384&100603&0.5&1&1    \\
\enddata
\tablecomments{
Observing log and properties of the 31 dpAGNs with close companions.
Column 1: J2000 designation. type-1s are indicated by stars.
Column 2: redshift.
Column 3: velocity splitting between \OIIIexpanded\ components.
Column 4: \citet{Kellermann89} radio loudness, $R = F_{\nu,\rm 5GHz}/F_{\nu,\rm 4400A}$.
Columns 5,6: Projected angular separation (arcsec) and physical
separation (kpc) between the main components in a merging system.
Column 7: name of the imager and the filter.
Column 8: total exposure time of the image.
Column 9: UT date (yymmdd) of the image.
Column 10: outside seeing at 5000\AA\ during the imaging observation.
Column 11: IFS data available? 1 - yes, 0 - no.
Column 12: source references---(1) \citet{Liu10a}, (2) \citet{Wang09}, (3) \citet{Smith10}.
}
\end{deluxetable}

\clearpage
\LongTables

\begin{deluxetable}{lccccccccc} 
\rotate
\tablewidth{0pt}
\tablecaption{Double-Peaked \OIII\ AGNs without Companions within 3\arcsec
\label{tab:imglog_single}}
\tablehead{ 
\colhead{SDSS Name} & \colhead{$z$} & \colhead{$\Delta$$V$} & \colhead{$R$} &
\colhead{Imager} & \colhead{Exptime} & \colhead{UT} & \colhead{Seeing} &
\colhead{IFS} & \colhead{Ref}\\
 & & (\kms) & & 
 & (second) & & (\arcsec) & 
 & \\
\colhead{(1)} & \colhead{(2)} & \colhead{(3)} & \colhead{(4)} & 
\colhead{(5)} & \colhead{(6)} & \colhead{(7)} & \colhead{(8)} &
\colhead{(9)} & \colhead{(10)}
}
\startdata
000249.1$+$004505 &0.087& 534&    7&   NIRC2/Kp& 192&110105&0.6&0&1,2  \\
000656.9$+$154848 &0.125& 382&  \nd&   NIRC2/Kp& 300&110105&0.7&0&2    \\
014209.0$-$005050 &0.133& 248&  \nd&   NIRC2/Kp& 240&110105&0.7&0&2    \\
040001.6$-$065254 &0.171& 391&   10& OSIRIS/Hbb& 240&100306&0.7&1&1    \\
073509.5$+$403624 &0.103& 264&  \nd&   NIRC2/Kp& 420&110105&0.7&0&2    \\
073656.5$+$475947 &0.096& 257&   21&   NIRC2/Kp& 240&110105&0.7&0&1    \\
073849.8$+$315612 &0.297& 297&  506&   NIRC2/Kp& 240&110105&0.7&0&1    \\
074129.7$+$392836 &0.210& 429&   28&   NIRC2/Kp& 240&110105&0.6&0&3    \\
074729.8$+$344018 &0.130& 319&  \nd&   NIRC2/Kp& 240&110105&0.6&0&2    \\
080741.0$+$390015 &0.023& 637&    2&WFPC2/F606W& 500&950226&\nd&0&2    \\
080841.2$+$481352 &0.124& 396&  \nd&   NIRC2/Kp& 720&110106&0.6&1&1,2,3\\
082107.9$+$502116 &0.095& 338&    6&   NIRC2/Kp& 360&110106&0.9&0&1,2  \\
082357.8$+$391631*&0.166& 580&  \nd&   NIRC2/Kp& 420&110105&0.5&0&3    \\
084130.2$+$393119 &0.132& 332&  \nd&   NIRC2/Kp& 480&110106&1.0&0&1    \\
084227.5$+$054716 &0.275& 490&  404&   NIRC2/Kp& 480&110106&0.8&0&1    \\
085358.6$+$233911 &0.113& 215&  \nd&   NIRC2/Kp& 480&110106&0.9&0&1    \\
090754.0$+$141509 &0.120& 285&  \nd&   NIRC2/Kp& 900&110106&0.8&0&1    \\
090947.9$+$312444*&0.264& 977&    2&   NIRC2/Kp& 420&110105&0.5&0&3    \\
090958.3$+$085542 &0.159& 341&   17&   NIRC2/Kp& 720&110106&0.9&0&3    \\
091654.1$+$521723*&0.219& 349&  \nd&   NIRC2/Kp& 448&110105&0.5&0&3    \\
094100.8$+$143614 &0.383& 854&   59&   NIRC2/Kp& 480&110106&1.2&0&3    \\
094124.0$+$394442 &0.108& 285& 2814&WFPC2/F555W& 600&990426&\nd&0&2    \\
094144.8$+$575124*&0.159& 339&  322&   NIRC2/Kp& 420&110105&0.5&0&3    \\
095833.2$-$005119 &0.086& 407&    3&   NIRC2/Kp& 480&110106&1.2&0&1,2  \\
095834.0$+$251235 &0.116& 336&    4&   NIRC2/Kp& 480&110106&0.9&0&1    \\
095920.5$+$422142 &0.153& 292&  \nd&   NIRC2/Kp& 480&110106&0.8&0&1    \\
100145.3$+$283330 &0.116& 513&  \nd&   NIRC2/Kp& 360&110106&0.8&0&3    \\
100654.2$+$464717 &0.123& 303&    4&   NIRC2/Kp& 480&110106&0.8&0&1    \\
101241.2$+$215556*&0.111& 257&  \nd&   NIRC2/Kp& 420&110105&0.5&0&3    \\
102004.4$+$324343*&0.484& 381&  \nd&   NIRC2/Kp& 420&110105&0.5&0&3    \\
103138.7$+$380652*&0.492& 734&  \nd&   NIRC2/Kp& 600&110105&0.5&0&3    \\
105052.5$+$083935 &0.169& 351&  \nd&   NIRC2/Kp& 480&110106&0.9&1&1    \\
111013.2$+$053339*&0.152& 755&  \nd&   NIRC2/Kp& 600&110105&0.4&0&3    \\
112659.5$+$294443 &0.102& 319&  \nd&   NIRC2/Kp& \nd&110106&0.9&0&1,2  \\
112907.1$+$575605 &0.313& 314&   36&   NIRC2/Kp& 540&110106&0.9&0&1    \\
113021.0$+$022212*&0.241& 402&  \nd&   NIRC2/Kp& 640&110105&0.4&0&3    \\
113045.3$+$505509*&0.592& 252&  \nd&   NIRC2/Kp& 420&110105&0.4&0&3    \\
113105.1$+$610405*&0.338& 427&  \nd&   NIRC2/Kp& 600&110105&0.4&0&3    \\
113257.8$+$604654*&0.233& 353&  \nd&   NIRC2/Kp& 420&110105&0.3&0&3    \\
114852.7$+$151416*&0.113& 306&  \nd&   NIRC2/Kp& 420&110105&0.4&0&3    \\
114908.5$+$144547*&0.595& 208&  \nd&   NIRC2/Kp& 420&110105&0.4&0&3    \\
115249.3$+$190300 &0.097& 323&    9&   NIRC2/Kp& 448&100603&0.6&0&1,2  \\
120240.7$+$263139*&0.476& 466&   66&   NIRC2/Kp& 192&100604&0.4&0&3    \\
120320.7$+$131931 &0.058& 292&  156&   NIRC2/Kp& 224&100603&0.6&0&2    \\
121911.2$+$042906*&0.555& 459&  \nd&   NIRC2/Kp&  64&100603&0.7&0&3    \\
121957.5$+$190003 &0.117& 390&  \nd&   NIRC2/Kp& 448&100603&0.6&0&1    \\
131106.7$+$195234 &0.156& 314&  \nd&   NIRC2/Kp& 448&100603&0.7&0&1    \\
131236.0$+$500416 &0.116& 353&   16&   NIRC2/Kp& 128&100604&0.4&0&2    \\
132104.6$-$001446 &0.082& 330&  \nd&   NIRC2/Kp& 448&100603&0.7&0&1    \\
140231.6$+$021546 &0.180& 276& 2553&   NIRC2/Kp& 448&100603&0.7&0&1    \\
140534.8$+$244735 &0.130& 218&    3&   NIRC2/Kp& 224&100604&0.4&0&1    \\
140845.7$+$353218 &0.166& 371&   16&   NIRC2/Kp& 448&100603&0.7&0&1    \\
144012.8$+$615633*&0.275& 505&    2&   NIRC2/Kp& 320&100603&0.7&1&3    \\
144157.2$+$094859 &0.220& 826&   61&   NIRC2/Kp& 320&100604&0.4&0&3    \\
145156.8$+$301603 &0.158& 512&  \nd&   NIRC2/Kp& 320&100603&0.8&0&1    \\
150437.7$+$541150*&0.305& 719&  \nd&   NIRC2/Kp& 256&100603&0.7&1&3    \\
150452.3$+$321415 &0.113& 250&    7&   NIRC2/Kp& 384&100603&0.8&0&2    \\
151659.2$+$051752 &0.051& 326&    8&   NIRC2/Kp& 160&100604&0.5&0&1,2  \\
151757.4$+$114453 &0.227& 433&  \nd&   NIRC2/Kp& 320&100603&0.7&0&3    \\
152606.2$+$414014 &0.008& 281&   30&WFPC2/F606W& 500&940828&\nd&0&2    \\
153231.8$+$420343*&0.209& 353&  \nd&   NIRC2/Kp& 448&100604&0.4&1&3    \\
155009.6$+$080839 &0.232& 400&   23&   NIRC2/Kp& 128&100604&0.3&0&1    \\
160027.8$+$083743 &0.227& 422& 3765&   NIRC2/Kp& 320&100604&0.3&0&1    \\
160436.2$+$500958 &0.146& 366&  \nd&   NIRC2/Kp& 320&100604&0.3&0&1,2  \\
160631.4$+$273643 &0.158& 319&    9&   NIRC2/Kp& 128&100604&0.4&0&1    \\
161826.9$+$081951*&0.446& 384&  137&   NIRC2/Kp& 160&100603&0.6&1&3    \\
161847.9$+$215925*&0.334& 365&    5&   NIRC2/Kp& 160&100603&0.6&1&3    \\
163056.8$+$164957 &0.034& 301&    2&   NIRC2/Kp& 192&100604&0.5&0&1    \\
163316.0$+$262716 &0.071& 307&  \nd&   NIRC2/Kp& 320&100603&0.6&0&2    \\
171544.0$+$600835 &0.157& 348&   62&   NIRC2/Kp& 320&100603&0.6&0&1,3  \\
171930.6$+$293413*&0.180&1105&    2&   NIRC2/Kp& 224&100603&0.6&1&3    \\
230442.8$-$093345 &0.032& 320&    8&   NIRC2/Kp& 320&100603&0.7&0&1,2  \\
231052.0$-$090012 &0.094& 329&  138&  ACS/F606W& 720&050819&\nd&0&1,2  \\
233313.2$+$004912 &0.170& 434& 1234&   NIRC2/Kp& 384&100603&0.7&1&1    \\
235256.6$+$001155 &0.167& 322&  \nd&   NIRC2/Kp& 384&100604&0.5&1&1    \\
\enddata
\tablecomments{
Observing log and properties of the 75 dpAGNs without close companions in \hst\ or
Keck/LGSAO images.
Column 1: J2000 designation. type-1s are indicated by stars.
Column 2: redshift.
Column 3: velocity splitting between \OIIIexpanded\ components.
Column 4: \citet{Kellermann89} radio loudness, $R = F_{\nu,\rm 5GHz}/F_{\nu,\rm 4400A}$.
Column 5: name of the imager and the filter.
Column 6: total exposure time of the image.
Column 7: UT date (yymmdd) of the image.
Column 8: outside seeing at 5000\AA\ during the imaging observation.
Column 9: IFS data available? 1 - yes, 0 - no.
Column 10: source references---(1) \citet{Liu10a}, (2) \citet{Wang09}, (3) \citet{Smith10}.
}
\end{deluxetable}

\end{document}